\documentclass[%
 aip,
 amsmath,amssymb,
 reprint,%
author-year,%
]{revtex4-1}

\usepackage{graphicx}
\usepackage{dcolumn}
\usepackage{bm}

\usepackage[utf8]{inputenc}
\usepackage[T1]{fontenc}
\usepackage{mathptmx}
\usepackage{etoolbox}

\usepackage{color}
\usepackage{soul}
\usepackage{xfrac}

\newcommand{\clara}[2]{\st{#1}{\color{black}{#2}}}

\makeatletter
\def\@email#1#2{%
 \endgroup
 \patchcmd{\titleblock@produce}
  {\frontmatter@RRAPformat}
  {\frontmatter@RRAPformat{\produce@RRAP{*#1\href{mailto:#2}{#2}}}\frontmatter@RRAPformat}
  {}{}
}%
\makeatother
\begin{document}
\renewcommand{\arraystretch}{2}

\preprint{AIP/123-QED}

\title[Triad interactions investigated by dual vortex shedding]{Triad interactions investigated by dual vortex shedding}
    
\author{Preben Buchhave}
    \affiliation{Intarsia Optics, S{\o}nderskovvej 3, 3460 Birkerød, Denmark.}

\author{Mengjia Ren}
    \affiliation{Department of Aeronautics and Astronautics, Kyushu University, 744 Motooka, Nishi-ku, Fukuoka 819-0395, Japan}

\author{Clara M. Velte}%
    \email{cmve@dtu.dk}
\affiliation{Department of Civil and Mechanical Engineering, Technical University of Denmark, Nils Koppels Allé, Building 403, 2800 Kongens Lyngby, Denmark.
}%

\date{\today}


\begin{abstract}
    The study of the exchange of momentum and energy between wave components of the turbulent velocity field, the so-called triad interactions, offers a unique way of visualizing and describing turbulence. Most often, this study has been carried out by Direct Numerical Simulations or by power spectral measurements. Due to the complexity of the problem and the great range of velocity scales in high Reynolds number developed turbulence, direct measurements of the interaction between the individual wave components have been rare. In the present work, we therefore present measurements and related computer simulations of triad interactions between controlled wave components injected into an approximately laminar and uniform flow from an open wind tunnel by vortex shedding from two rods suspended into the flow. \clara{}{This well-defined vortex shedding approximates well a two-dimensional flow, which makes the analysis of the triadic interactions considerably less complex to analyze than in a fully developed spatially three-dimensional flow. With the information obtained from the simulations, we are hereby able to isolate and display the individual triad interactions taking place as the flow develops downstream as well as the strengths of these interactions. The experiments provide the time constants governing the development of higher order frequency components. The combination of these experiments and simulations} provide unique insight into the inner workings of turbulence and shows how the nonlinear term in the Navier-Stokes equation \clara{}{on average} forces the energy towards higher frequencies, which is the reason for the so-called energy cascade.
\end{abstract}


\maketitle

\section{Introduction}

Turbulence is characterized by a spatial mixture of fluid velocity fluctuations spanning a great range of frequencies. The dynamics of these fluctuations is described mathematically by the Navier-Stokes equation (NSE), and the growth and interchange between them are governed by the nonlinear term in the NSE. This nonlinearity is of a special form, involving the product of a vector quantity, the velocity, and its own derivative, the velocity spatial gradient. This second order nonlinearity generates new frequency components that are phase matched to either sum or differences of two existing frequency components, motivating the term triad interactions. Because of the special form of the nonlinearity, the coefficient for the sum frequency is greater than the one for the difference frequency (see e.g.~\cite{NSMachine}), which explains the tendency of the spectral energy to transfer towards higher frequencies, the so-called energy cascade. 

Because of the great range of sizes of the spatial velocity fluctuations and the very different roles they play in the interactions, whether it be forcing terms delivering energy to the flow, it be fluctuations in an equilibrium state of the turbulence or whether it concerns small fluctuations losing energy to viscous diffusion, it is exceedingly difficult, due to the combination of strong nonlinearity and the effect of dissipation, to give a detailed accurate description of the dynamics of turbulence. However, a successful study of the triad interactions offers an instructive insight into the inner ``machinery'' of turbulence.

Triad interactions have in the past most often been analyzed analytically by means of Fourier integrals resulting in delta function phase match conditions where the resulting wave vector, say $\vec{k}$, must equal exactly the vector sum of two incident Fourier components with wave vectors $\vec{k}_1$ and $\vec{k}_2$ giving the phase match condition $\vec{k} = \vec{k}_1 + \vec{k}_2$. The existence of temporal and spatial windows, which are encountered in all practical measurements, implies that the interaction between Fourier components is allowed outside the strict delta-function phase match condition, for example within a $\mathrm{sinc}^2$-shaped spectral window in case of a box car window in the physical domain. A detailed study of triad interactions as they appear in real experiments is treated in~\cite{DynamicTriadInts} (see also~\cite{ClaraiTi}).

The study of triad interactions in Fourier space offers a unique way of viewing the basic energy exchange between velocity scales taking place when a turbulent flow develops and how spatial velocity structures, in a net sense, develop towards smaller and smaller scales. In realistic, practical flows, the complexity is so great that it is difficult to follow the individual interactions. Thus, there is a need to devise experiments where fewer Fourier components are introduced into a flow and the development of these components and the interactions between them and the surrounding flow can be isolated and traced as the flow evolves. However, not many such experiments are to be found in the literature. An earlier study by our group describes measurements of vortex shedding behind a single rectangular rod inserted across a low turbulence uniform flow and behind oscillating airfoils in a highly turbulent flow,~\cite{Dotti2020}. By spectral analysis, it was possible to follow the development of the oscillating modes introduced in the flow. Other similar studies, focused mainly on the aerodynamic properties of airfoils, show indirectly some of the same spectral development, but without analyzing the flow in terms of triad interactions, see e.g.~\cite{Wuetal2004} and~\cite{JUNG2005451}.

In earlier work attempting to explain the spectral transfer, different models for the coupling between the Fourier components describing the spectral distribution of the fluctuations have been proposed, c.f.~\cite{batchelor1953theory} and~\cite{Heisenberg1948}. Best known and most used for deriving average properties of flows at high Reynolds numbers is probably the theory initiated by~\cite{Kolmogorov1941}, often referred to as `K41', which postulates that the small-scale fluctuations in flows at very high Reynolds numbers are essentially decoupled from the large scales that are infusing the energy into the flow, and that the fluctuations in the medium and high frequency range are in statistical equilibrium. It then follows that the interactions between the small, high frequency scales and the large energy containing scales is simply a convection of the small scales by the large scales without any direct coupling through the nonlinear term in the NSE (although the NSE in itself does not exclude such interactions). The spectral energy transfer then operates as a cascade between similar size spectral Fourier components only governed by the rate of energy input by the large energy containing scales and the energy is dissipated predominantly at the small scales. Dimensional analysis then consequently explains many properties of equilibrium turbulence. 

However, none of the early proposed theories for spectral transfer derives directly from the widely accepted governing equation of motion for turbulence -- the NSE. Recent experimental evidence, as also pointed out early by~\cite{kraichnan_1959}, hints at the weakness of the Kolmogorov picture and its inability to describe non-equilibrium turbulence, see also~\cite{George2013} and~\cite{ClaraiTi}. A significant problem with K41 is that it does not take into account, for example, strong interactions between very disparate scales as caused for example by large-scale stretching of small-scale vortex filaments and vortex sheets,~\cite{Kraichnan1974}. Neither are isolated occurrences of high activity velocity bursts considered~(\cite{BnT1949}), although a later expansion of the K41 theory attempts to include these effects,~\cite{Kolmogorov1962}. 

Also missing in the K41 theory is an explanation of the persistence of large-scale structures throughout developing flows like jets and wakes, c.f.~\cite{yeung_brasseur_wang_1995}. Direct numerical simulations of the NSE has provided helpful insight into the interactions between Fourier modes and the spectral transfer, but the substantial computational costs still limit the calculations to relatively low Reynolds numbers.

The question about local vs. nonlocal interactions and the role played by the large structures have been an issue since early on. \cite{Batchelor1949}, for example, showed theoretically that the large scale structures, determined by the initial conditions, survived until well into the process of decay of the fine structures in homogeneous turbulence, indicating a weak coupling between the small and large eddies. An experimental demonstration of this can be found in  e.g.~\cite{Dotti2020}. 

\cite{YeungBrasseur1991},~\cite{yeung_brasseur_wang_1995} and~\cite{brasseurwang1992} conducted early studies by Direct Numerical Simulations (DNS) of the interaction between the large structures and the anisotropy of the small structures and showed that the small scales were influenced by nonlocal interactions between the low frequency/wavenumber Fourier components (large scales, low spatial frequency) and two closely spaced high frequency Fourier components (small scales). Thus, many studies contradict the simpler Kolmogorov picture of independence between large and small structures and a cascade dominated by interactions between neighboring modes. Further DNS studies of local vs. nonlocal interactions have been reported by~\cite{Domaradzki1990}. Their results show that energy transfer is predominantly taking place between scales of similar size, but that the triad interactions are caused by separate, nonlocal wave vectors. 

Triad interactions in turbulence have been studied indirectly theoretically by modelling the process (see~\cite{Kraichnan1962} for some early work), experimentally by power spectral measurements~(\cite{van_atta_chen_1969}) and numerically by DNS simulations (c.f.~\cite{Domaradzki1990} and~\cite{Ohkitani1992}). \clara{}{A recent study by means of two-dimensional particle image velocimetry of a turbulent flow behind an array of rectangular rods of different dimensions (\cite{BajandBuxton2017}) was able to distinguish both location and spectrum of the triadic production terms generated by nonlinear combinations of the multi-scale turbulence shed from the rods. Although impressive by managing to present both spatial location and strength of the spectral components, this work does not, as in our case, show explicitly which exact triad interactions take part in the generation of the power spectrum.}

In the following, we describe measurements similar to those of~\cite{Dotti2020}, but with two rectangular rods instead of a single one. The objective being to study combinations of deterministic low frequency components and how they interact locally and/or nonlocally as the flow evolves. \clara{}{The simple two-dimensional design of the experiment is motivated by the fact that it allows to isolate and simplify the triadic interactions, since the wavenumber vectors can only point in two directions along the same vertical axis (i.e., upwards or downwards). Fully developed spatially three-dimensional turbulence of course obeys the same physics (i.e., equations), but the two-dimensional experiment simplifies the analysis by reducing the dimensionality of the problem. Furthermore, it isolates the observed interactions to the ones resulting from the originally generated frequencies and thereby considerably simplifies the analysis.} 

\clara{}{Comparing the measured spectra to the computer simulations allows us to identify exactly which frequency combinations take part in the triad interactions, quantify the energy exchanges between the components and in which sequence they occur. Although the measurements and interpretations are made in an approximately two-dimensional flow, the results can of course be directly be extended to three-dimensional flows, but at the cost of considerable increased complexity.} Corresponding computations with an NSE-simulator, applied and validated against measurements in~\cite{NSMachine}, agree well with the measurement results and identify the individual triad interactions, their strengths and time constants and provide details of the energy exchange between components.

\section{Background theory}

\subsection{Navier-Stokes equation in Fourier space}

The motion of an incompressible fluid with constant kinematic viscosity at a small test volume at position $\vec{r} = (x,y,z)$ and time $t$ is described in physical space by mass conservation (\ref{eqn:1}) and momentum conservation (\ref{eqn:2}):
\begin{equation}\label{eqn:1}
    \nabla \cdot \vec{u}(\vec{r},t) = 0
\end{equation}
\begin{equation}\label{eqn:2}
    \frac{\partial \vec{u}(\vec{r},t)}{\partial t} + \vec{N}(\vec{r},t) = - \nabla p (\vec{r},t) + \nu \nabla^2 \vec{u}(\vec{r},t) 
\end{equation} 
Here $\vec{u} = \vec{u}(\vec{r},t)$ is the velocity at the test volume, $p = p(\vec{r},t)$ is pressure, $\nu$ is the kinematic viscosity and the density is normalized to unity, $\rho = 1$. The second term on the left hand side in the NSE (\ref{eqn:2}), the nonlinear convection term $\vec{N}(\vec{r},t) = \left ( \vec{u} (\vec{r},t)\cdot \nabla \right ) \vec{u} (\vec{r},t)$ of the NSE, makes this equation quite difficult to solve for fluids of high or even moderate Reynolds numbers, since it involves the product of the fluctuating velocity and the velocity gradient over a large range of frequencies. However, the term is local, meaning that it can be solved at each point separately. The third term, the pressure term, is nonlocal, since it involves the instantaneous sum of pressure fluctuations generated throughout the full flow field.

As mentioned, it may give a different, and perhaps better, insight into the momentum and energy transfer between different spatial scales of the flow if we consider the NSE in Fourier space \clara{}{for a homogeneous flow}. The four-dimensional Fourier transform, \clara{}{designated by the circonflex} $\hat{\cdot}$, of the velocity is given by
\begin{equation}
    \hat{\vec{u}}(\vec{k},\omega ) = \int_{-\infty}^{\infty} \int_{-\infty}^{\infty} \int_{-\infty}^{\infty} \int_{-\infty}^{\infty} \vec{W}(\vec{r}) W(t) e^{-i (\vec{k} \cdot \vec{r} - \omega t)} \vec{u} (\vec{r},t) \, d\vec{r} \, dt
\end{equation}
where $\omega$ is the temporal frequency. The window functions $\vec{W}(\vec{r})$ and $W(t)$ reflect the fact that any real experiment is of limited spatial and temporal extent.

Mass conservation and momentum conservation in Fourier space is then given by equations (\ref{eqn:3}) and (\ref{eqn:4}) (\cite{DynamicTriadInts}):
\begin{equation}\label{eqn:3}
    \vec{k} \cdot \hat{\vec{u}}(\vec{k},\omega) = 0
\end{equation}
\begin{equation}\label{eqn:4}
    - i \omega \hat{\vec{u}}(\vec{k},\omega) + \left [ 1 - \frac{\vec{k}}{k^2}\vec{k}\cdot \right ] \hat{\vec{N}}(\vec{k},\omega) = -\nu k^2 \hat{\vec{u}}(\vec{k},\omega)
\end{equation}
The first term in the square bracket is due to the convection term and the second one is due to the pressure gradient term. Thus, the nonlinear effects can be studied by considering only the Fourier transform of the nonlinear term, $\hat{\vec{N}}(\vec{k}, \omega)$. This Fourier term may be expressed by:
$$\hat{\vec{N}}(\vec{k}, \omega) = \frac{1}{(2 \pi)^4} \iiiint \, d\vec{k}_1\, d\omega_1  \, \, \frac{1}{(2 \pi)^4} \iiiint \, d\vec{k}_2\, d\omega_2$$
$$\times \iiiint \vec{W}(\vec{r}) W(t) e^{-i \left [ (\vec{k} - \vec{k}_1 - \vec{k}_2 )\cdot \vec{r} - (\omega - \omega_1 - \omega_2)t \right ] } \, d\vec{r} \, dt $$
\begin{equation}
\times \left [\left ( i\vec{k}_2 \cdot \hat{\vec{u}} (\vec{k}_1,\omega _1)\right ) \hat{\vec{u}} (\vec{k}_2 , \omega_2)\right ]
\end{equation}

In a highly turbulent flow, the mode interactions involve all four-dimensional phase match conditions, $\left [ (\vec{k} - \vec{k}_1 - \vec{k}_2 )\cdot \vec{r} - (\omega - \omega_1 - \omega_2)t \right ]=0$. However, \clara{}{since we can to a good approximation in our experiments invoke Taylor's hypothesis (the velocity fluctuations are $\sim 2\%$ of the freestream velocity)}, the four-dimensional mode match simplifies to the usual three-dimensional triad condition, $\vec{k} = \vec{k}_1 + \vec{k}_2$, see~\cite{DynamicTriadInts}.

However, even a strong vortex, as generated in this experiment, will contain a range of wavenumbers due to the effect of the finite window functions. As explained below, the wave vectors for the strong vortices generated in the experiment all point in the same \clara{}{or opposite} direction, and the phase match condition then becomes a scalar condition, $k = k_1 + k_2$. Using Taylor's hypothesis \clara{}{along the downstream direction}, we can equally express the mode match condition by the measured temporal frequencies, $\omega = \omega_1 + \omega_2$ or $f = f_1 + f_2$, where $\omega = 2 \pi f = u_0 k$ and $u_0$ is the local convection velocity. 

The interactions between spatial frequency components can be analyzed by a simple example as described in~\cite{NSMachine}. If Taylor's hypothesis can be invoked, the analogous analysis can be written in terms of temporal frequencies, $f=u_0\frac{k}{2 \pi}$. Considering two parallel spatial velocity waves with different wavenumbers, $\vec{k}_1$ and $\vec{k}_2$ (since the wavenumber vectors are parallel, the direction is indicated only by the sign in the below):
$$
u(s) = u_1 \cos \left (k_1 s \right ) + u_2 \cos \left (k_2 s \right )
$$
where $s$ is a spatial coordinate along the convected flow through the measurement point,~\cite{buchhave2017measurement} \clara{}{and $u_1$ and $u_2$ are the amplitudes of the respective velocity waves}. When inserted into the nonlinear term:
$$
u(s) \frac{d}{ds} u(s) = 
$$
$$
\left [ \left ( u_1 \cos \left (k_1 s \right ) + u_2 \cos \left (k_2 s \right ) \right ) \cdot \frac{d}{ds} \left ( u_1 \cos \left (k_1 s \right ) + u_2 \cos \left (k_2 s \right ) \right ) \right ]
$$
$$
= - \left [ \left ( u_1 \cos \left (k_1 s \right ) + u_2 \cos \left (k_2 s \right ) \right ) \cdot \left ( k_1 u_1 \sin \left (k_1 s \right ) + k_2 u_2 \sin \left (k_2 s \right ) \right ) \right ]
$$
$$
= -\frac{1}{2} [ k_1 u_1^2 \sin \left (2k_1 s \right ) + k_2 u_2^2 \sin \left (2k_2 s\right )
$$
\begin{equation}\label{eq:HSexample}
+(k_1+k_2) u_1 u_2 \sin [(k_1+k_2)s] +(k_1-k_2) u_1 u_2 \sin [(k_1-k_2)s]  ]
\end{equation}
\clara{}{Setting $u_1 = u_2 = 1$, we obtain the same result as in~\cite{NSMachine}.} We see that four frequencies will be generated, doubling of the two original frequencies, sum and difference of the two frequencies. The sum of the two original frequencies will have the largest coefficient, especially when $k_1=k_2$, and the energy will have a \clara{}{net} tendency towards higher frequency values. \clara{}{Note that Equation (7) only presents a single product term between oscillations that may represent two single modes of the velocity field.}

The two cases of single and dual rod vortex shedding will now be analyzed based on these insights. 

\subsection{Triad interactions for a single rectangular rod}

Rectangular rods have been used in the experiments, since they are able to generate narrow spectral peaks from shedding, in contrast to cylindrical rods. When a rectangular rod is placed in a laminar flow, vortices will be generated behind it by the process of vortex shedding. Close to the rod, only a single vortex appears with a single, fundamental frequency. In the wave vector space, this fundamental vortex can be represented by the wave vector $\vec{k}_1$\clara{}{, pointing alternately upwards and downwards in the direction parallel to the rod if the two-dimensional flow assumption is applied.} At this time, there is only one scale in the vortex shedding if the incoming flow is assumed to be completely laminar, and the further development will only happen because of the nonlinear convection of this wave vector. 

The first new wave vector to appear will be $2\vec{k}_1$, generated by the process of frequency doubling as deducible from~(\ref{eq:HSexample}). The frequency difference will in this case generate a zero frequency contribution. When the $2\vec{k}_1$ is formed, further combinations are possible; One is $2\vec{k}_1$ interacting with $\vec{k}_1$ to produce $3\vec{k}_1$ (frequency tripling) through frequency summation. Another one is frequency quadrupling of $2\vec{k}_1$ (or summing a wavenumber with itself), which generates $4\vec{k}_1$. Frequency subtraction by e.g. $2k_1-k_1=k_1$ also contributes, however with a much weaker interaction, since the coefficient, $(2k_1 - k_1)$, is smaller than for the other two. 

\begin{figure}[!h]
    \includegraphics[width=0.85 \linewidth]{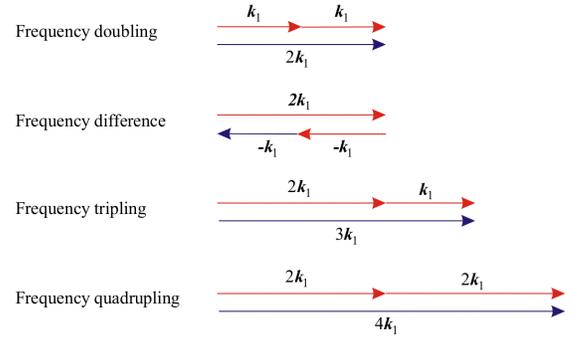}\\
    \caption{Basic initial interactions in the wake of a single rod. Red contributions correspond to the initial components and blue contributions to the predicted resulting frequency from the non-linear triad interactions, see (\ref{eq:HSexample}).}
    \label{fig:singlerod}
\end{figure}

Some of the processes are shown in Figure~\ref{fig:singlerod}, where the red contributions correspond to the initial components and blue contributions to the predicted resulting frequency from the (parallel) non-linear triad interactions, see (\ref{eq:HSexample}). The mentioned combinations belong only to the basic interactions at the very beginning of the development of the velocity field. Further downstream, more scales will be formed, and more possible combinations will appear. 

\subsection{Triad interactions for dual rectangular rods}

\begin{figure}[!h]
    \includegraphics[width=0.85 \linewidth]{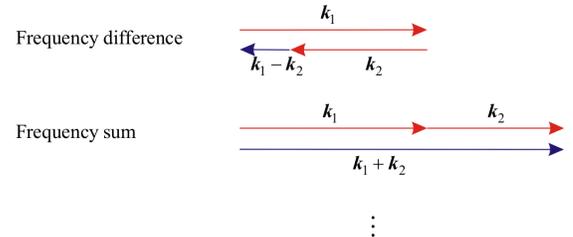}\\
    \caption{Basic initial interactions in the wake of dual rods. Red contributions correspond to the initial components and blue contributions to the predicted resulting frequency from the (parallel) non-linear triad interactions, see (\ref{eq:HSexample}).}
    \label{fig:dualrod}
\end{figure}

When positioning two rectangular rods in parallel in a laminar flow, each rod will shed vortices, and the two vortices and their individual developments will interact with each other downstream. Initially, the possible scales in the wake will be the same as discussed for the single rod. However, further downstream the vortices will interact with each other and more wave vector interaction combinations will be possible as compared to the single rod case. If the fundamental wave vectors generated by the two rods are denoted $\vec{k}_1$ and $\vec{k}_2$, then downstream the flow will contain integer multiples of both $\vec{k}_1$ and $\vec{k}_2$ as well as sums and differences of all possible combinations of $\vec{k}_1$ and $\vec{k}_2$. A schematic diagram of some basic initial interactions is shown in Figure~\ref{fig:dualrod}. \clara{}{As the wave vectors of the two vortex streets are pointing alternately up and down, we see that there is a symmetry in the process such that the sum of two vectors pointing in the same direction corresponds to the difference when one of the vectors changes direction. The power spectrum, being the square of the Fourier components (and also averaged over time), will not be affected by the alternating vector directions.}

\section{\label{sec:Measurements}Method}

\subsection{Measurement Method}

\begin{figure}[!h]
    \includegraphics[width=0.85 \linewidth]{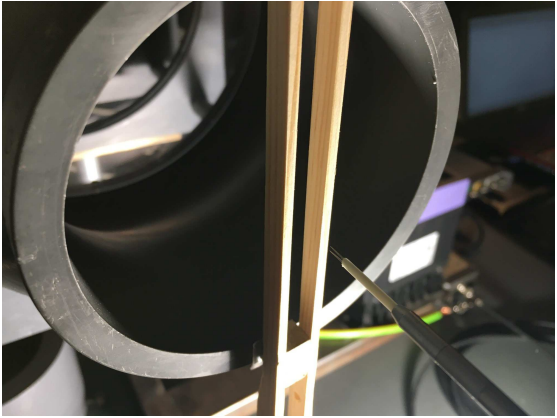}\\
    \caption{Two rectangular rods mounted vertically across the $100\, mm$ diameter jet exit.}
    \label{fig:1}
\end{figure}

Vortices were generated by respectively one or two rectangular rods vertically suspended near the center diameter of an open wind tunnel (in the form of a round jet in air). Figure~\ref{fig:1} displays an example of one of the dual-rod setups. The measurements were performed in the near field of the $100\, mm$ diameter open wind tunnel with contraction ratio of 2.4:1. The inside of the nozzle follows a 5th order polynomial, ensuring a nearly laminar top-hat velocity profile at the jet exit~(c.f.~\cite{wygnanski_fiedler_1969} and \cite{hussein_capp_george_1994}). The flow was conditioned by internal screens to a turbulence intensity level of less than 1\% at the jet exit. The Reynolds number based on the exit diameter and exit velocity was $Re=1650$ \clara{}{(based on the rectangular rod dimensions and jet exit velocity, $Re=40-50$)}. 

From previous experiments, we have noticed that rods with a rectangular cross section with sharp corners generate nearly sinusoidal signals with a narrow spectrum, permitting us to consider the vortices as singular oscillating modes injected into the main flow,~\cite{Dotti2020}. Moreover, the geometry of the setup allows us to well approximate these modes to be two-dimensional with velocity fluctuations mainly in a horizontal plane through the centerline of the jet and thus a wave vector for the mode in the \clara{}{(alternating upwards and downwards)} vertical direction. 

The frequency content of the flow was measured with a hot-wire probe placed with the wire oriented in the vertical direction (to reduce spatial filtering along the directions of the largest spatial variations) and mounted on a traversing system located far downstream of the jet exit. The probe used for measurements was a \clara{}{DANTEC} 55P11 straight single wire probe, which is made of tungsten, with a diameter of $5\,\mu m$ and length $1\, mm$. The probe was connected through a straight 55H20 support using a $4\, m$ long cable. The hot-wire was controlled by a DANTEC Mini CTA 54T30 system and the CTA signal was digitized by a Series 5000 Picoscope digital oscilloscope. 

The modulation of the velocity signal originating from the vortex shedding was only of the order of a few percent of the freestream velocity. It was thus not deemed necessary to linearize the CTA output. The record length was selected to $1\, s$ and the sampling frequency was chosen to $50\, kS/s$. The measured time records were stored for later further processing. The sampled signals were also processed directly on the Picoscope to provide power spectra ensemble averaged over a selected number of measured records. Since the raw voltage signals from the hotwire were used, the resulting power spectra are provided in dimensions `$dBu$'. The measurement positions were chosen based on where the vortex shedding is most pure and detectable (as could be directly monitored using the Picoscope).

\subsection{Computer Simulations -- Software description}

The simulations are based on an iterative one-dimensional solution of the full Navier-Stokes equation (NSE) including all the terms, see~\cite{NSMachine}. The forces acting on a small flow control volume at a fixed Eulerian measurement point in the flow are projected onto the direction of the instantaneous velocity vector in the control volume. This one-dimensional record contains information about the spatial structures convected through the test volume. Input into the algorithm is a one-dimensional digital finite velocity time record stemming from e.g. an analytical signal, a computer simulation or from a real measurement of the velocity at the measurement point with an Eulerian instrument, e.g. a hot-wire anemometer or a laser Doppler anemometer. 

The program computes the time development of the instantaneous velocity projected onto the instantaneous flow direction. This means that we can compute the time evolution of the full magnitude of the momentum, the kinetic energy as well as other quantities such as various structure functions, correlations and spectral quantities. We can e.g. also compute the time development of the full velocity power spectrum. As explained below, the computations involve summation over all the spatial Fourier components of the record, which means that we have access to all the triad interactions, when and how they develop and the respective strengths of all contributions to the spectrum.

The algorithm includes all the terms in the Navier-Stokes equation. However, as the pressure term is not local at the measurement point, but includes pressure fluctuations generated over the full flow field, including the pressure should ideally require the solution over the whole flow field. This is not possible with our algorithm, which includes only terms evaluated within the control volume. To include pressure, we have to rely on assumptions or models. As we deal here with few, large and energetic eddies, we can safely disregard the influence of the pressure term. \clara{}{Inserting values derived from the experiment, a rough estimate shows the pressure term being a factor of fifty smaller than the convective term. The reason is primarily the mean velocity being much greater than the fluctuation velocity of the vortex.} See also~\cite{NSMachine}, where pressure was included. \clara{}{Furthermore, since the pressure and convection terms enter the Navier-Stokes equation in the same manner, see e.g. equation~(\ref{eqn:4}), the nonlinear triad interactions will be unaltered except for perhaps the intensity.}

The most important features of the program are:
\begin{itemize}
    \item The calculations are based on a one-dimensional temporal velocity record, which is interpreted as the projection of the three-dimensional velocity vector onto the instantaneous flow direction at the measurement point.
    \item Knowing the instantaneous velocity magnitude,~\cite{buchhave2017measurement}, or applying Taylor’s hypothesis to the measured velocity time record if the turbulence intensity is below $\sim 20-25\%$, we can convert the velocity record to a spatial record and compute the spatial Fourier components.
    \item The input to the program can equally be a computer simulated signal with a few frequency components, a model for a turbulent flow with a given continuous spectrum or it can be a measured temporal velocity record.
    \item The program, dealing with only the one-dimensional components in the instantaneous flow direction, cannot give information on the spatial three-dimensional trajectory of the fluid particles.
\end{itemize}

\subsubsection{Block diagram of software}

\begin{figure}[!h]
    \includegraphics[width=0.85 \linewidth]{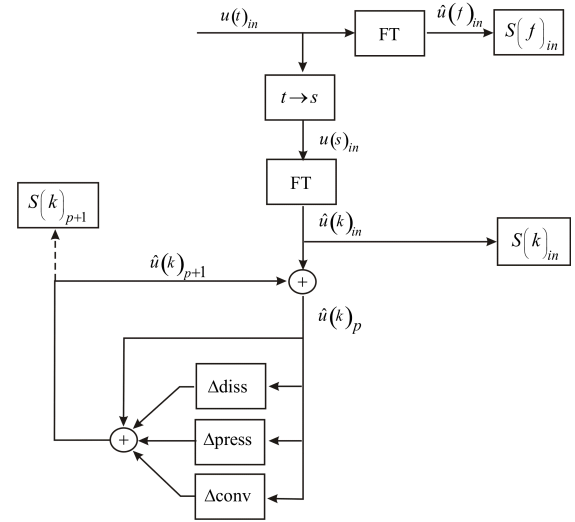}\\
    \caption{Block diagram of the Navier-Stokes simulation program, see~\cite{NSMachine}.}
    \label{fig:8}
\end{figure}

To conform with the Picoscope displays, we exhibit the spectra in the following in terms of temporal frequency. However, the algorithm computes the triad interactions in spatial Fourier space. The software algorithm is explained with reference to the block diagram shown in Figure~\ref{fig:8}. $u(t)_{in}$ is the input time record. This time record is Fourier transformed, $\hat{u}(f)_{in}$, and the input velocity power spectrum, $S(f)_{in}$, is displayed. The time record, $u(t)_{in}$, is also converted to a spatial velocity record, $u(s)_{in}$, where $s$ is the convection record length, using either the convection record method,~\cite{buchhave2017measurement}, or by using Taylor's hypothesis if the turbulence intensity is less than $\sim 20-25\%$. The spatial record is then Fourier transformed to provide the spatial input Fourier coefficients as a function of the wavenumber, $\hat{u}(k)_{in}$, and the spatial input power spectrum, $S(k)_{in}$, is displayed. The input signal is added to the circulating signal, $\hat{u}(k)_{p}$, where $p$ is the current iteration. The input signal can e.g. be added only in the first iteration, which would simulate a disturbance added to the flow at the beginning and then allowed to decay. Or, alternatively, a signal may e.g. be added at each iteration, corresponding to a forced flow. At each small time-increment, $\Delta t$, a small contribution from each force term in the NSE is added to the circulating field. As the contributions are considered infinitesimal, they can be added linearly in any order. The developing spatial power spectrum can be displayed after a selected number of iterations. Furthermore (not shown in the diagram), the growth of the Fourier components, real and imaginary parts, for selected wavenumbers can be displayed, allowing us to follow the development of the triad interactions as a function of time.

\section{Results}

Experiments \clara{}{and simulations} were performed with both a number of different single- and dual-rod geometries. The main difference resulting from different rod geometries is some difference in the stability of the vortex generation and the spectral purity of the vortex shedding signals. Since the purpose of the experiments was not a survey of vortex generation physics, but rather a study of the development of the vortices considered as single oscillatory modes, we present only results of the single- and dual-rod experiments with clean signals.   

\clara{}{We begin by comparing and discussing the single rod experiments and simulations, followed by a corresponding presentation of the dual rod results. }

\subsection{Single Rod Results}

\subsubsection{Experiments}\label{sec:singleresults}

\begin{figure}[!h]
    \includegraphics[width=0.85 \linewidth]{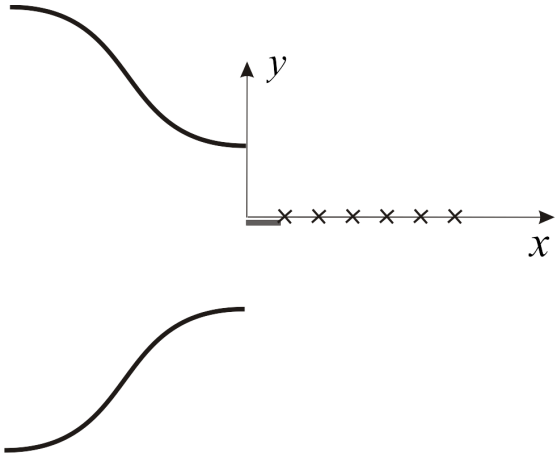}\\
    \caption{Top view of the single-rod geometry. Rod cross section: $10 \times 2 \, mm$. Measurement positions: $y=1.5\, mm$ and $x=10$, $20$, $30$, $40$, $50$ and $60 \, mm$.}
    \label{fig:2}
\end{figure}

A top view of the single-rod experiments geometry is shown in Figure~\ref{fig:2}. The geometry is, for practical purposes, described in a Cartesian coordinate system with the x-axis along the centerline of the jet and the y-axis in the horizontal transverse direction. The z-axis is positive in the vertical upward direction. 

The rod cross section was $10 \times 2 \, mm$. The rod was positioned symmetrically across the center of the jet at $y=0 \, mm$, and the hotwire measurement positions were chosen based on the quality of the measured vortex shedding signal, resulting in measurement positions $y=1.5\, mm$ and $x=10$, $20$, $30$, $40$, $50$ and $60 \, mm$ downstream from the jet orifice. 

\begin{figure}[!h]
    \includegraphics[width=0.85 \linewidth]{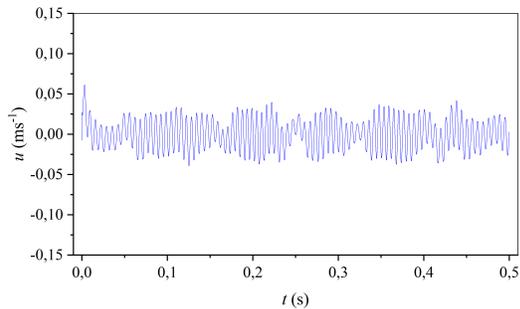}\\
    \caption{A $0.5\, s$ time trace of the velocity fluctuations at $x = 10 \, mm$, $y=1.5\, mm$ and $z=0\, mm$.}
    \label{fig:3}
\end{figure}

Figure~\ref{fig:3} shows the fluctuating part of the measured velocity signal near the back edge of the rod at $y=1.5\, mm$ and $x=10\, mm$. We cannot expect the signal to show a continuous long sinusoidal form. Vortex shedding will contain some irregular phase jumps due to the three-dimensionality of the geometry. In addition, low frequency fluctuations occur due to disturbances in the jet flow introduced from the surroundings. The signal shown is the AC-part of a much larger DC-signal. The velocity modulation due to the vortex shedding is approximately 2\% of the mean velocity. 

\begin{figure*}
\centering
\begin{minipage}{0.5\textwidth}
  \centering
  \includegraphics[width=0.99\linewidth]{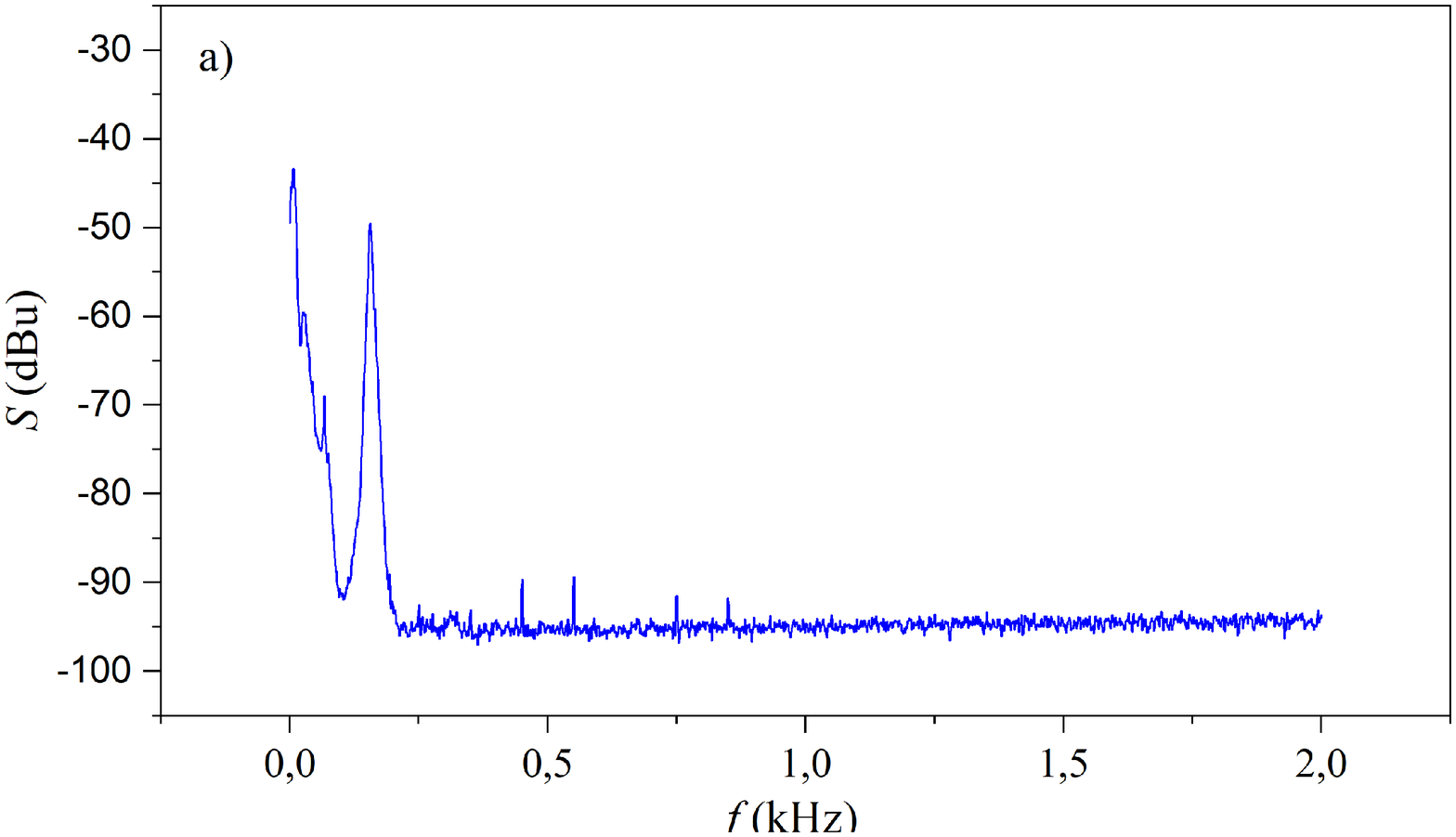}
\end{minipage}%
\begin{minipage}{.5\textwidth}
  \centering
  \includegraphics[width=0.98\linewidth]{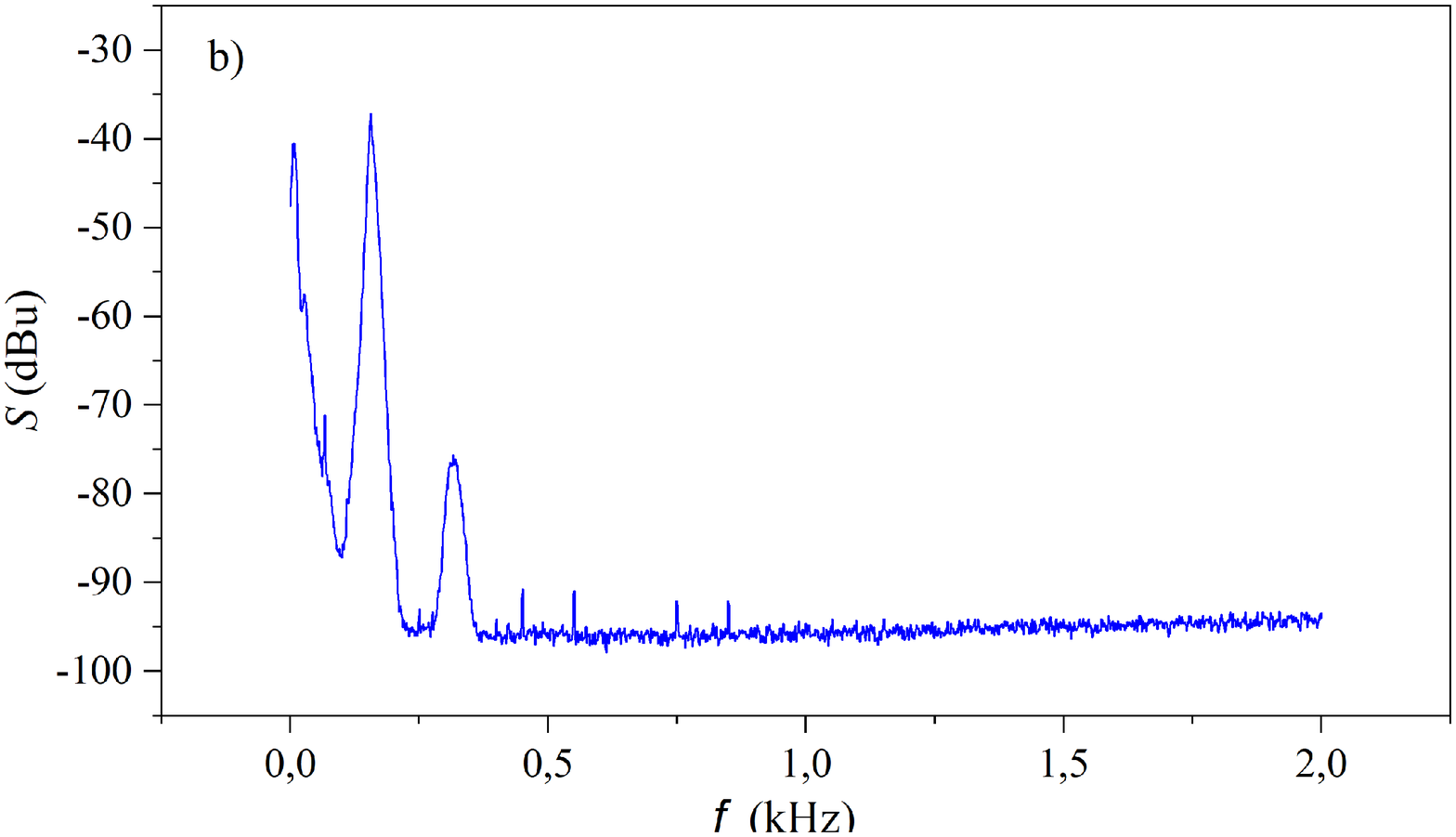}
\end{minipage}
\begin{minipage}{0.5\textwidth}
  \centering
  \includegraphics[width=0.99\linewidth]{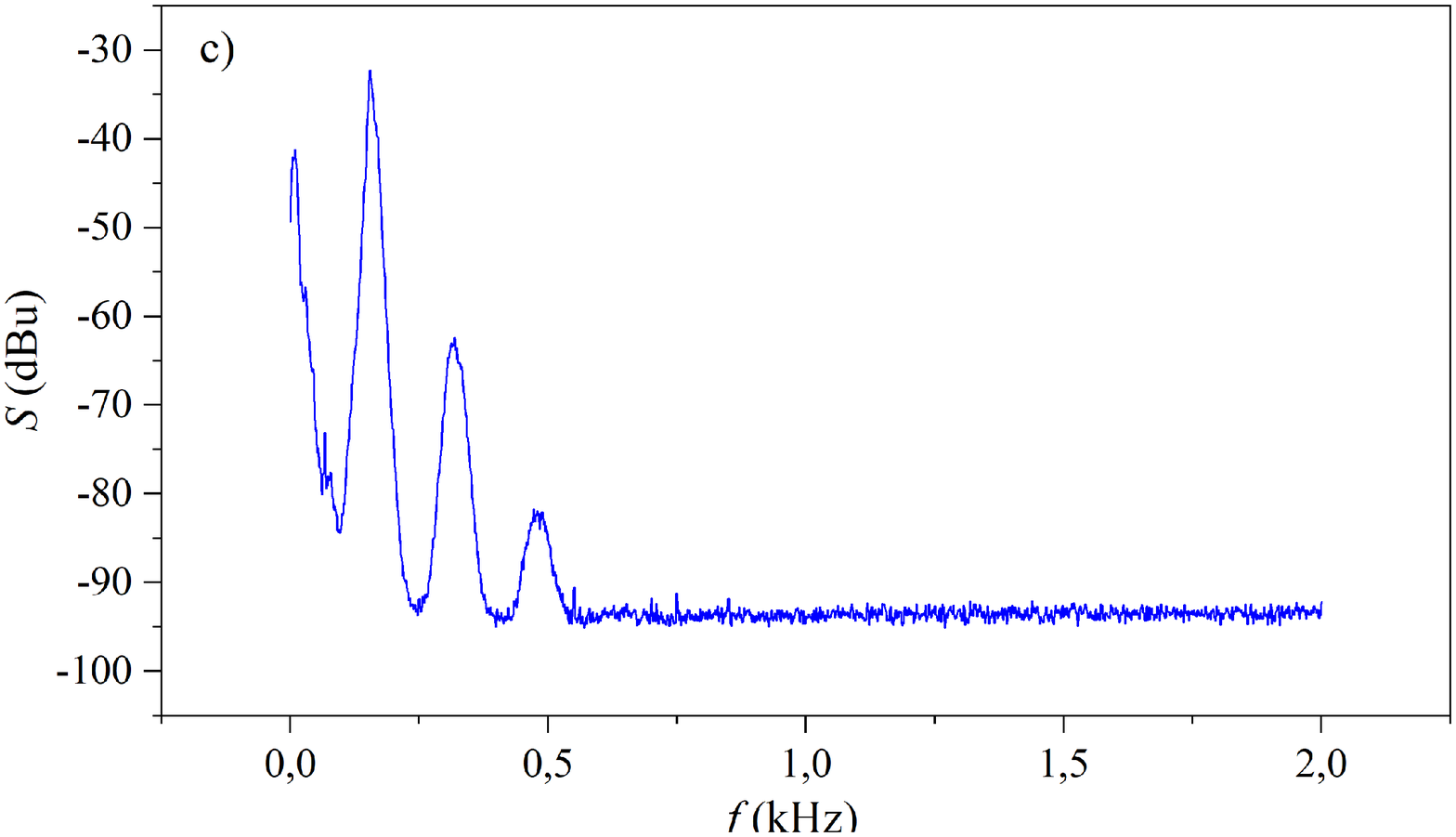}
\end{minipage}%
\begin{minipage}{.5\textwidth}
  \centering
  \includegraphics[width=0.98\linewidth]{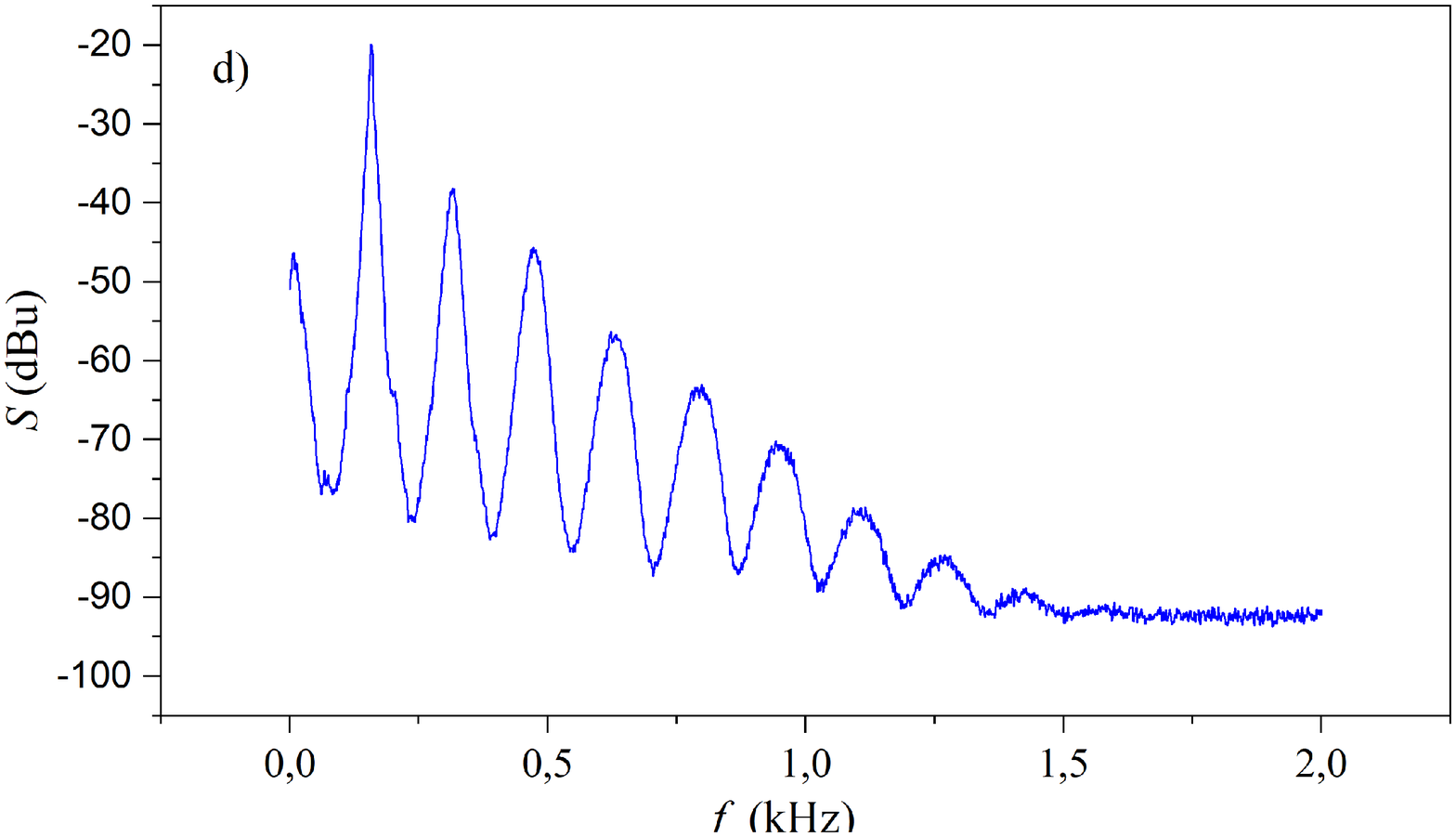}
\end{minipage}
\begin{minipage}{0.5\textwidth}
  \centering
  \includegraphics[width=0.99\linewidth]{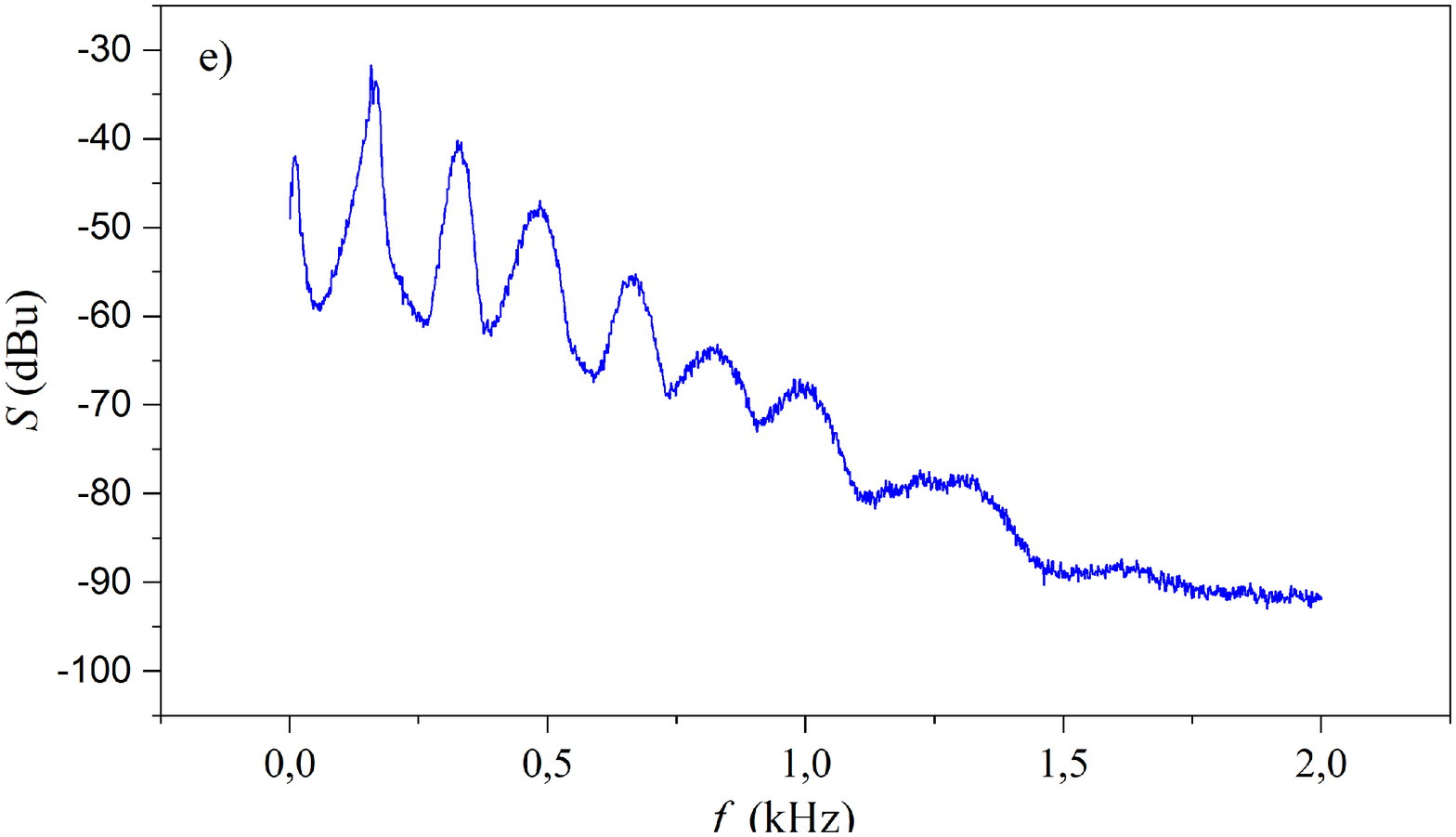}
\end{minipage}%
\begin{minipage}{.5\textwidth}
  \centering
  \includegraphics[width=0.98\linewidth]{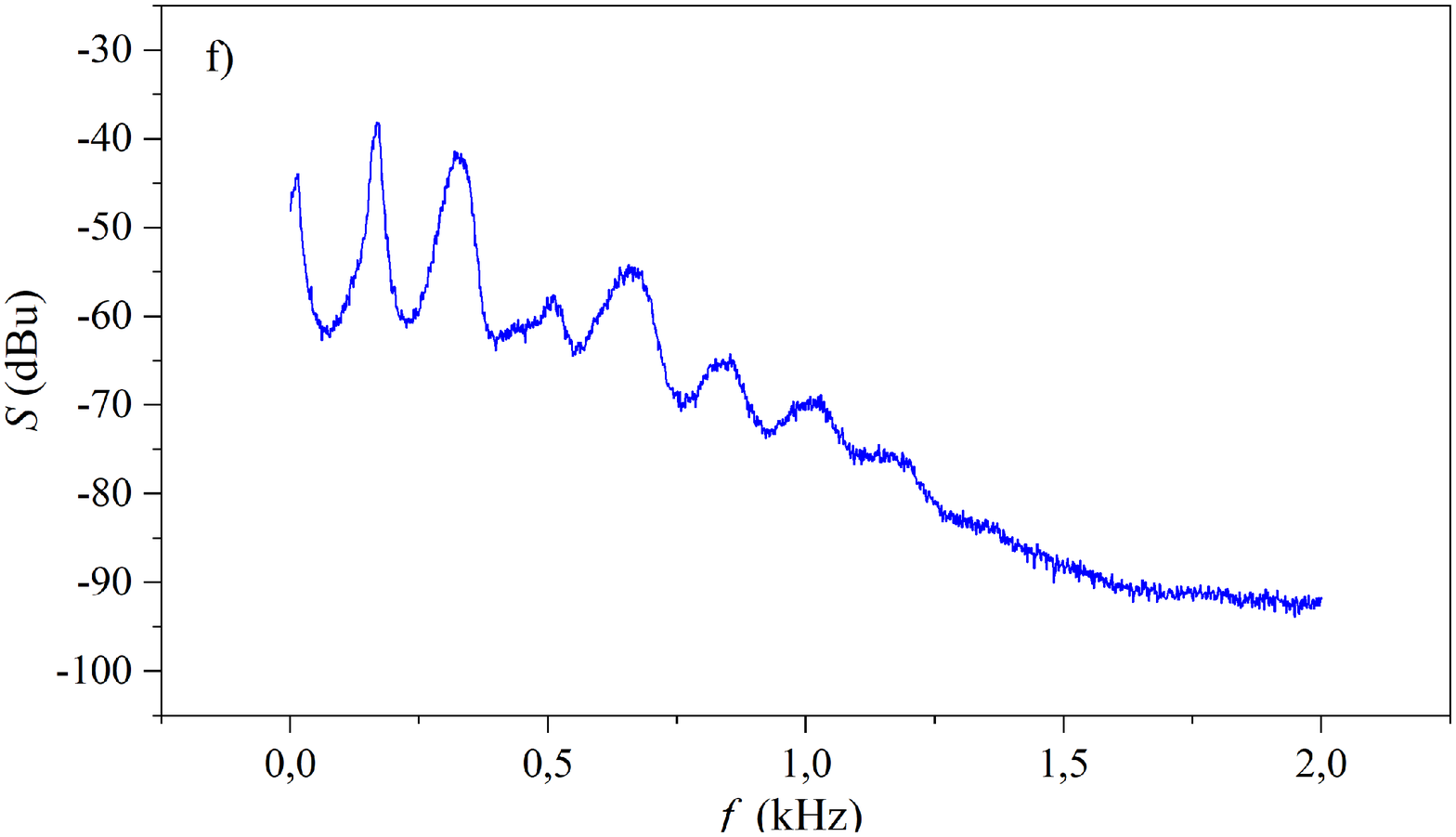}
\end{minipage}
    \caption{Time development of the measured single-rod power spectrum of the hotwire voltage at $y = 1.5\, mm$ and, respectively, at (a) $x= 10 \, mm$, (b) $x= 20 \, mm$,  (c) $x= 30 \, mm$,  (d) $x= 40 \, mm$,  (e) $x= 50 \, mm$ and  (f) $x= 60 \, mm$.}
    \label{fig:4}
\end{figure*}

Figures~\ref{fig:4}\textit{a-f} show, respectively, the development of the power spectrum, $S$ with units $(dBu)$, of the hotwire voltage signal at $y=1.5\, mm$ and $x=10$, $20$, $30$, $40$, $50$ and $60\, mm$. Already at $x=20\,mm$ we see the second harmonic being formed, and as we move downstream, we see an increase in the number of higher harmonics. Some of the peaks are second harmonics of peaks at lower frequency and some are the result of combinations of two different lower frequencies. In the present case, the fundamental vortex frequency is centered around $156\, Hz$. The second peak is clearly the second harmonic centered around $312\, Hz$. The third peak at $468\, Hz$ is the sum of the previous two.

For more downstream positions, as more and more peaks appear, there exist several possible combinations and the analysis becomes rapidly increasingly complex. The fourth peak could be the double of the second peak and/or the sum of the first and the third. Which ones we see and how strongly they enter cannot be deduced from studying the power spectrum, but in section~\ref{sec:singlesims} on computer simulations on the single rod case, we shall see the details of the different combinations, when they occur and with what strength they contribute. Figures~\ref{fig:4}\textit{e} and~\ref{fig:4}\textit{f} show that the regular series of distinct peaks begins to approach a continuous spectrum, which happens for several reasons; With downstream distance, the peaks increase in number and interact increasingly with the background turbulence from the merging shear layers and also lose power due to dissipation.

\subsubsection{Simulations}\label{sec:singlesims}

\begin{figure*}
\centering
\begin{minipage}{0.5\textwidth}
  \centering
  \includegraphics[width=0.99\linewidth]{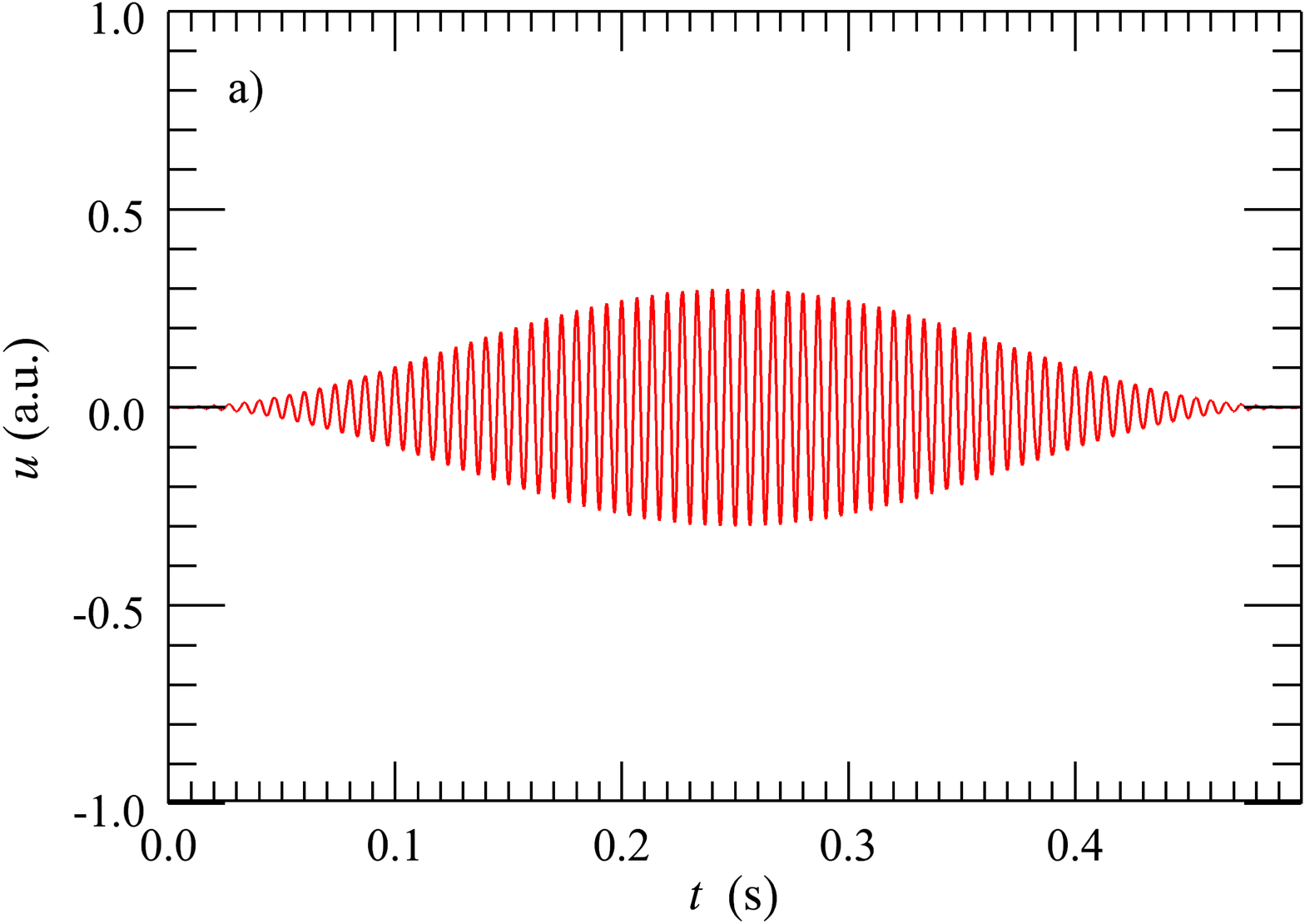}
\end{minipage}%
\begin{minipage}{.5\textwidth}
  \centering
  \includegraphics[width=0.98\linewidth]{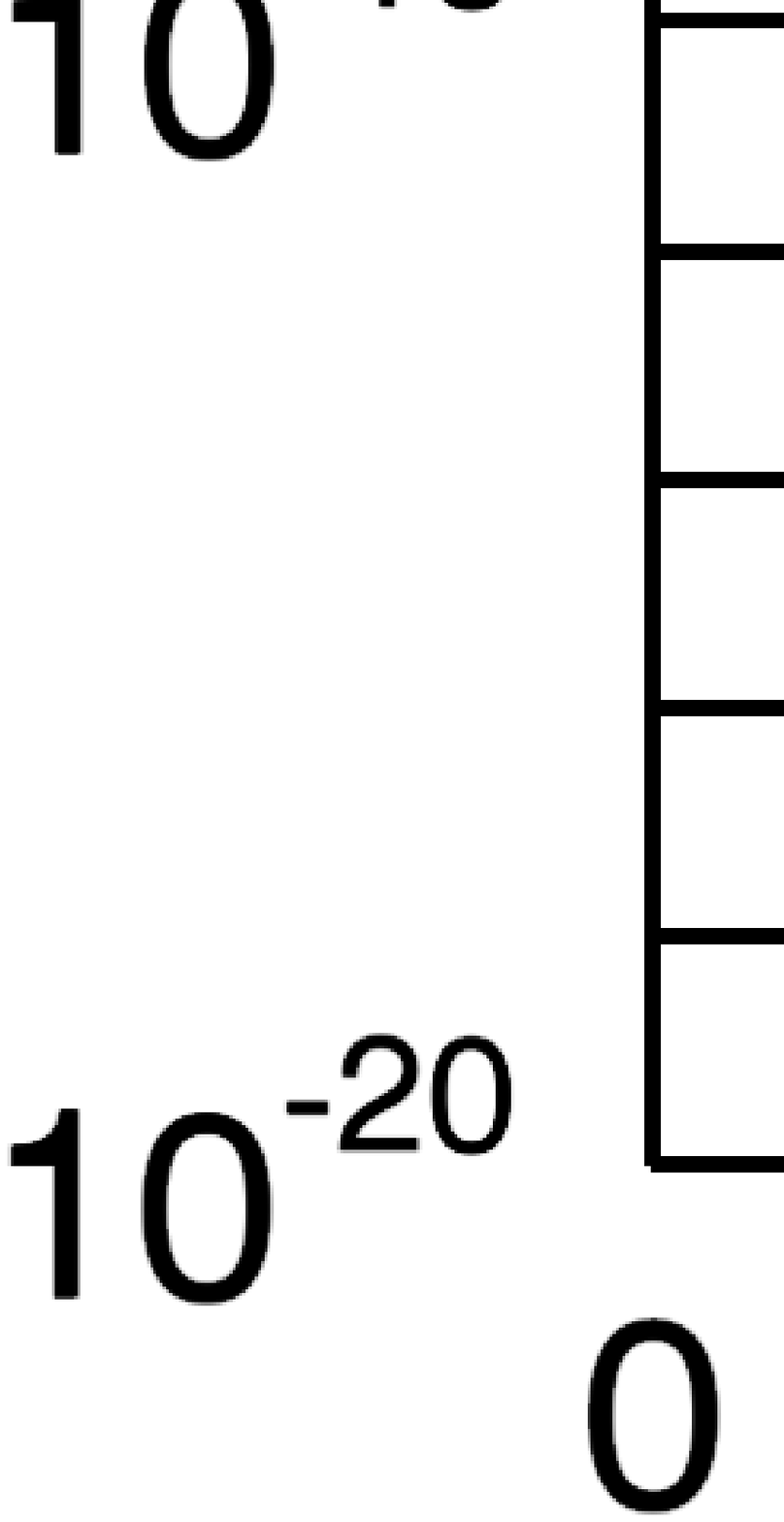}
\end{minipage}
    \caption{(a) A $0.5\, s$ time record (incident signal) of a Gaussian pulse with modulation frequency $150\, Hz$ and (b) the corresponding power spectrum.}
    \label{fig:9}
\end{figure*}

Figure~\ref{fig:9}\textit{a} shows the input signal used in the single rod simulations. The dimensions of the input signal (and all estimates derived therefrom) are set to arbitrary units, `a.u', since the simulations output a synthetic velocity. The time signal is a $0.5\, s$ record of a part of a Gaussian pulse modulated with a frequency of $f=150\, Hz$, meant to represent the vortex shedding frequencies close to the single rod in the experiment described in Section~\ref{sec:singleresults}. The corresponding power spectrum of this input signal is shown in Figure~\ref{fig:9}\textit{b}. 

\begin{figure*}
\centering
\begin{minipage}{0.5\textwidth}
  \centering
  \includegraphics[width=0.99\linewidth]{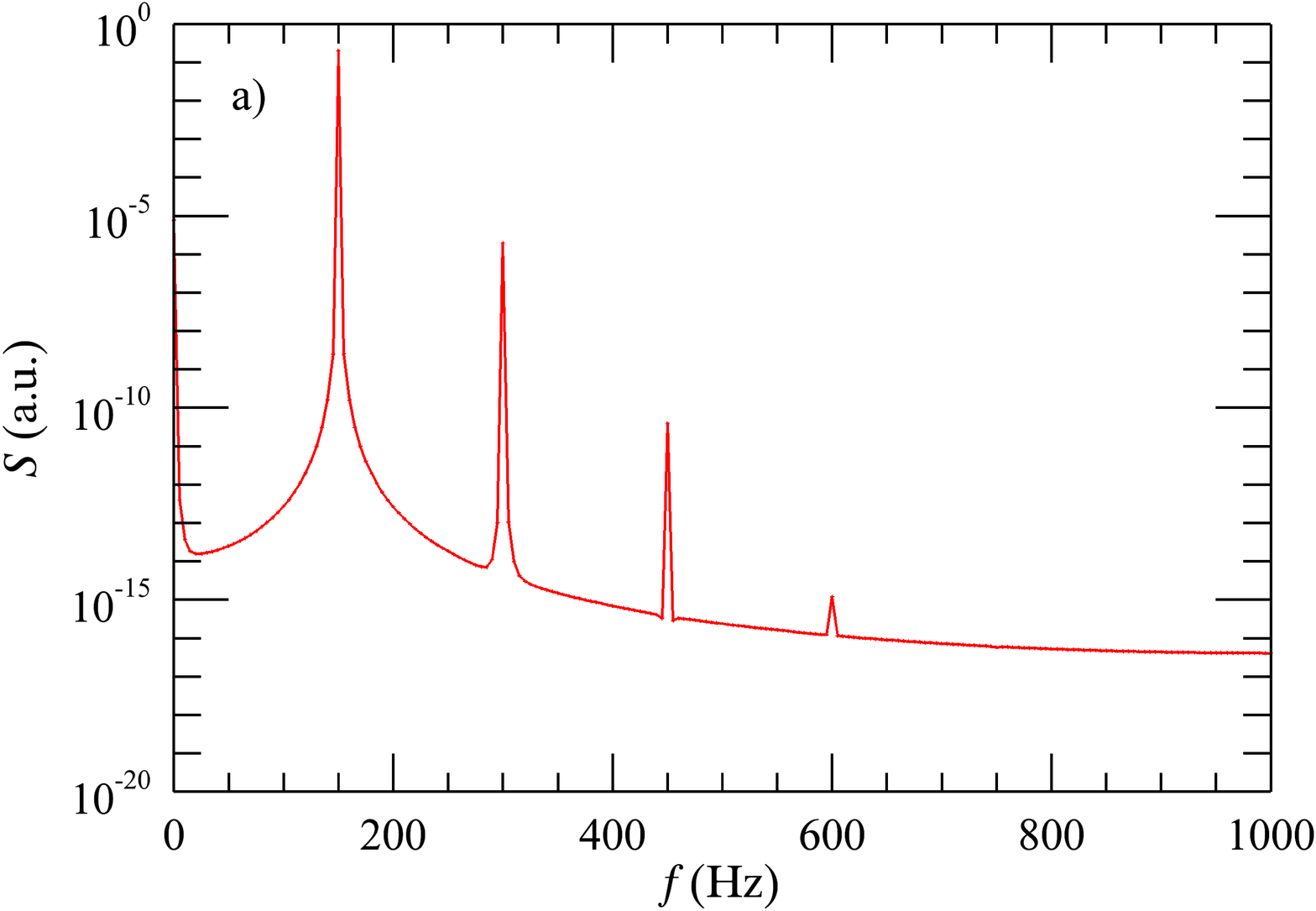}
\end{minipage}%
\begin{minipage}{.5\textwidth}
  \centering
  \includegraphics[width=0.98\linewidth]{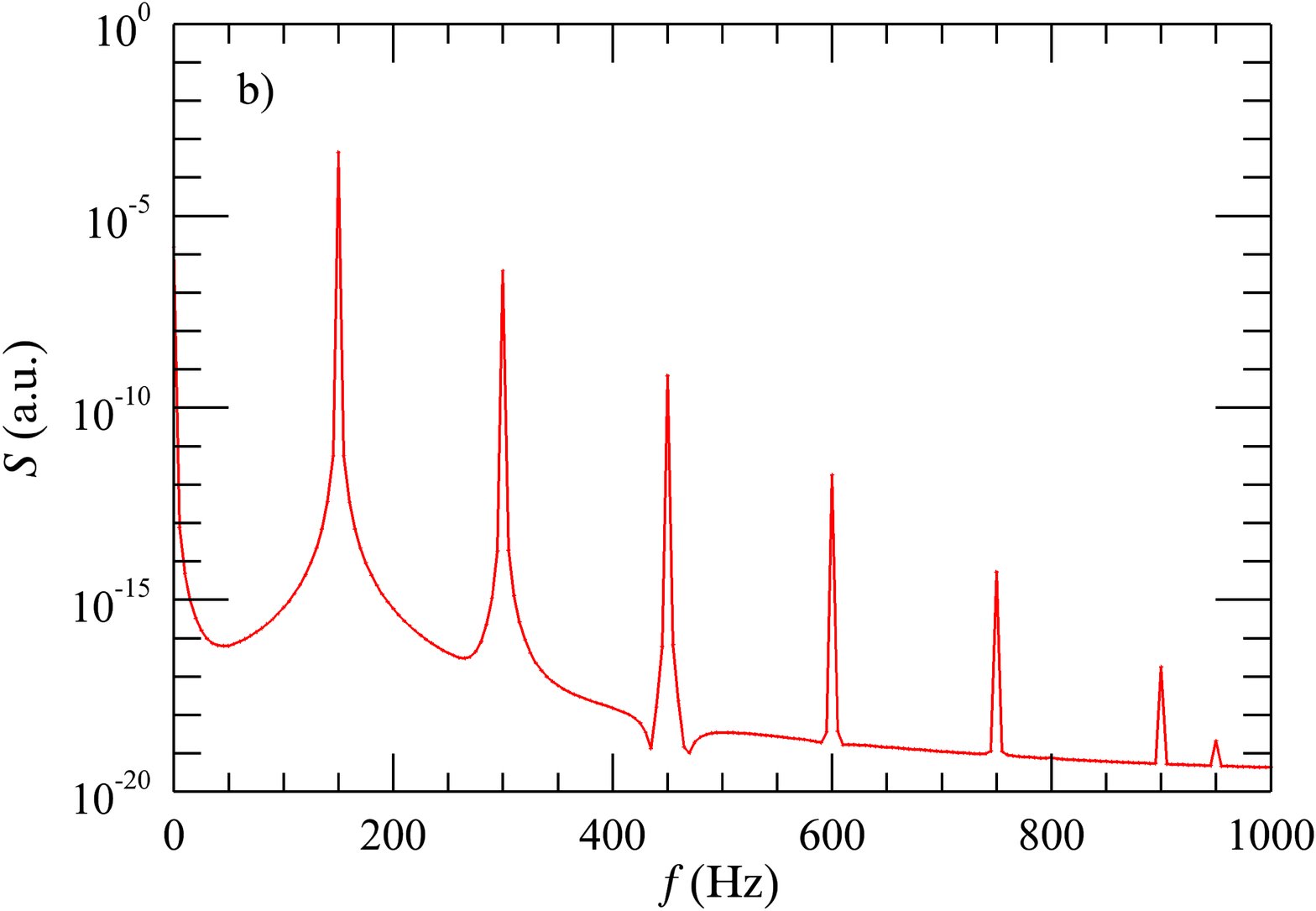}
\end{minipage}
    \caption{Time development of single frequency power spectrum. (a) 100 iterations, (b) 200 iterations.}
    \label{fig:10}
\end{figure*}

Figures~\ref{fig:10}\textit{a} and~\ref{fig:10}\textit{b} show the time development of this signal after 100 and 200 iterations, respectively. The peaks occur at the possible phase match values of $f$, $2f$, $4f$ and also at possible sum and difference combinations of two of these frequencies \clara{}{(including $f=0$\,Hz)}.


\begin{figure*}
\centering
\begin{minipage}{0.5\textwidth}
  \centering
  \includegraphics[width=0.99\linewidth]{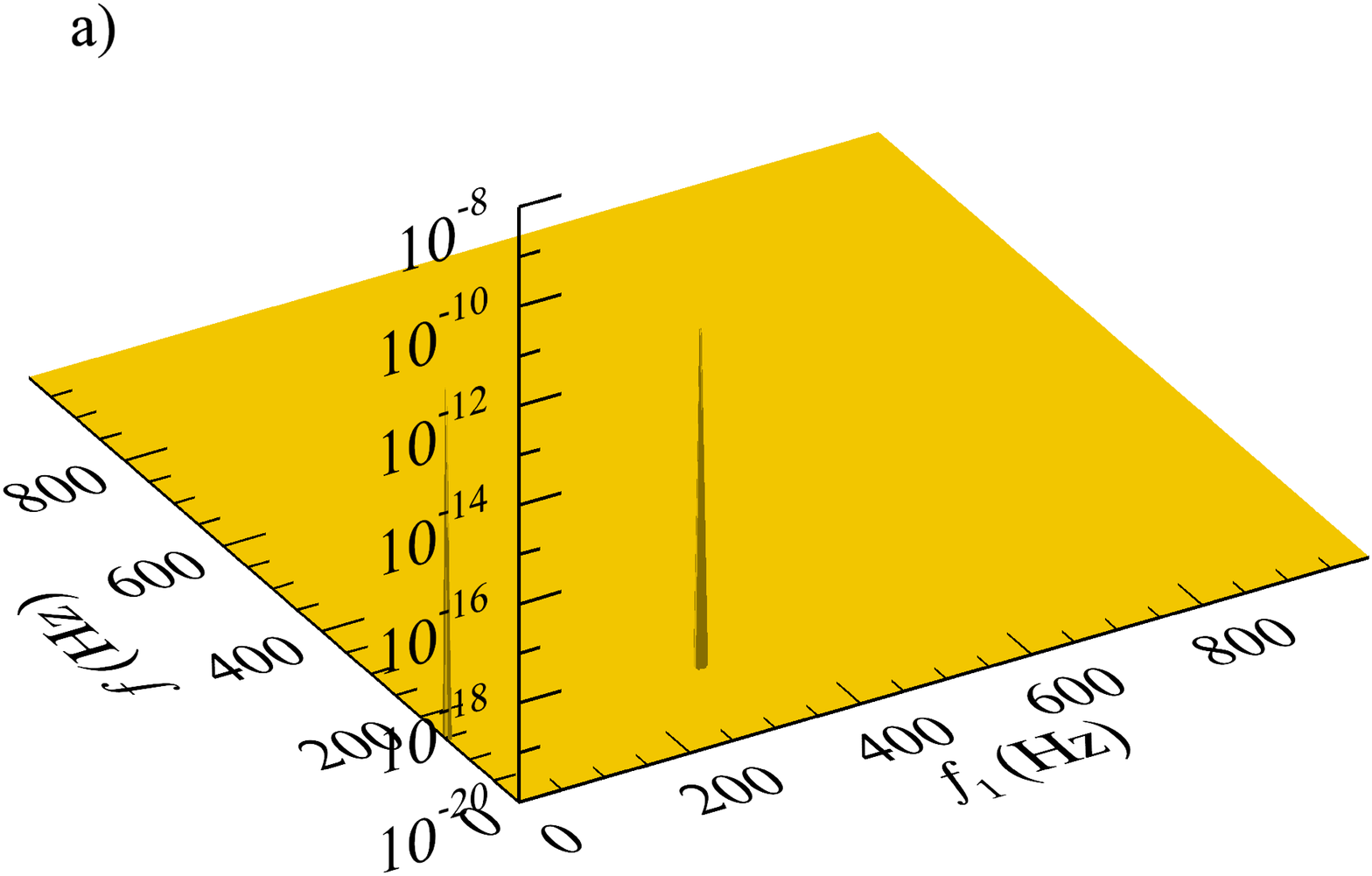}
\end{minipage}%
\begin{minipage}{.5\textwidth}
  \centering
  \includegraphics[width=0.98\linewidth]{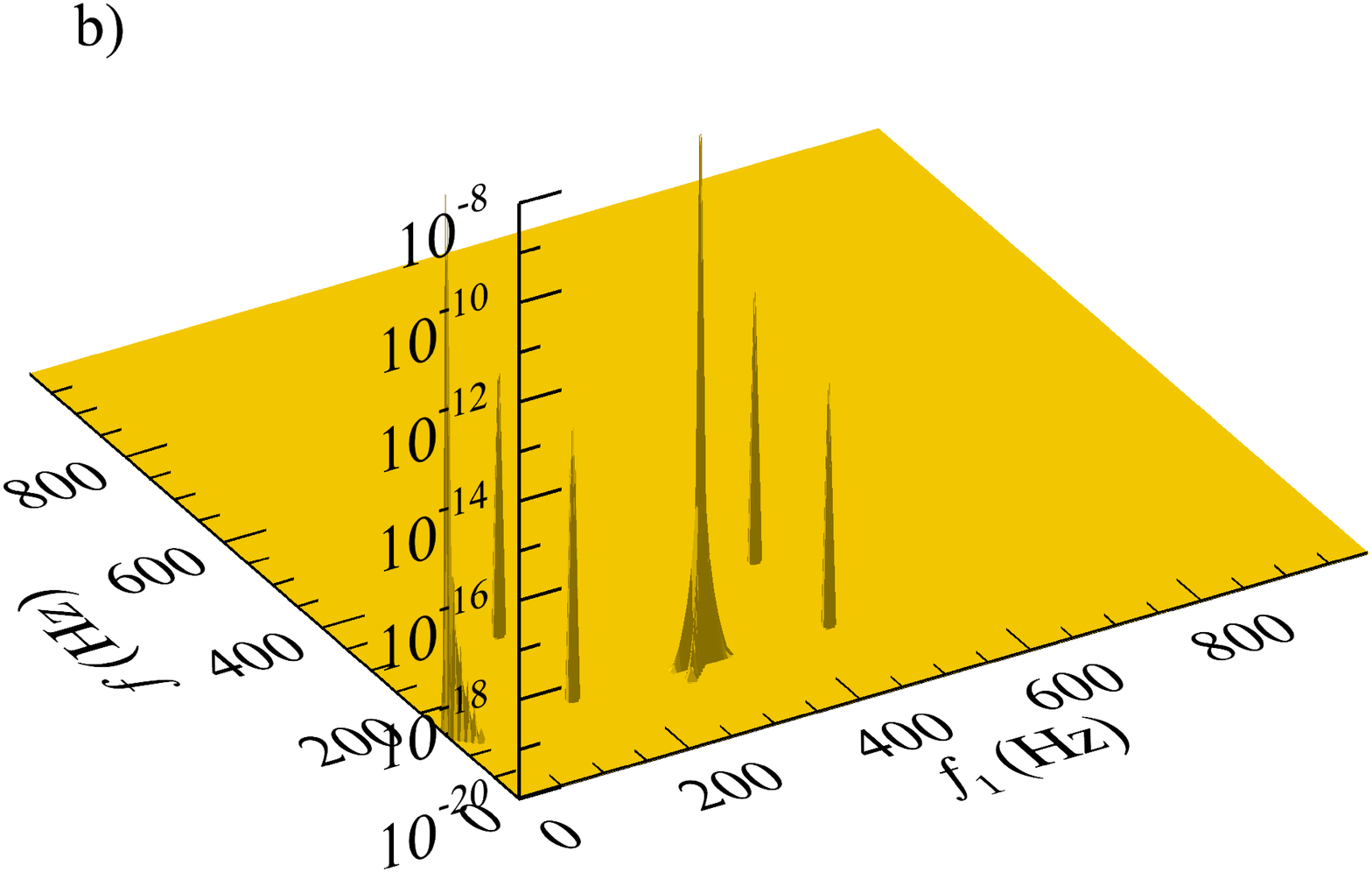}
\end{minipage}
\begin{minipage}{0.5\textwidth}
  \centering
  \includegraphics[width=0.99\linewidth]{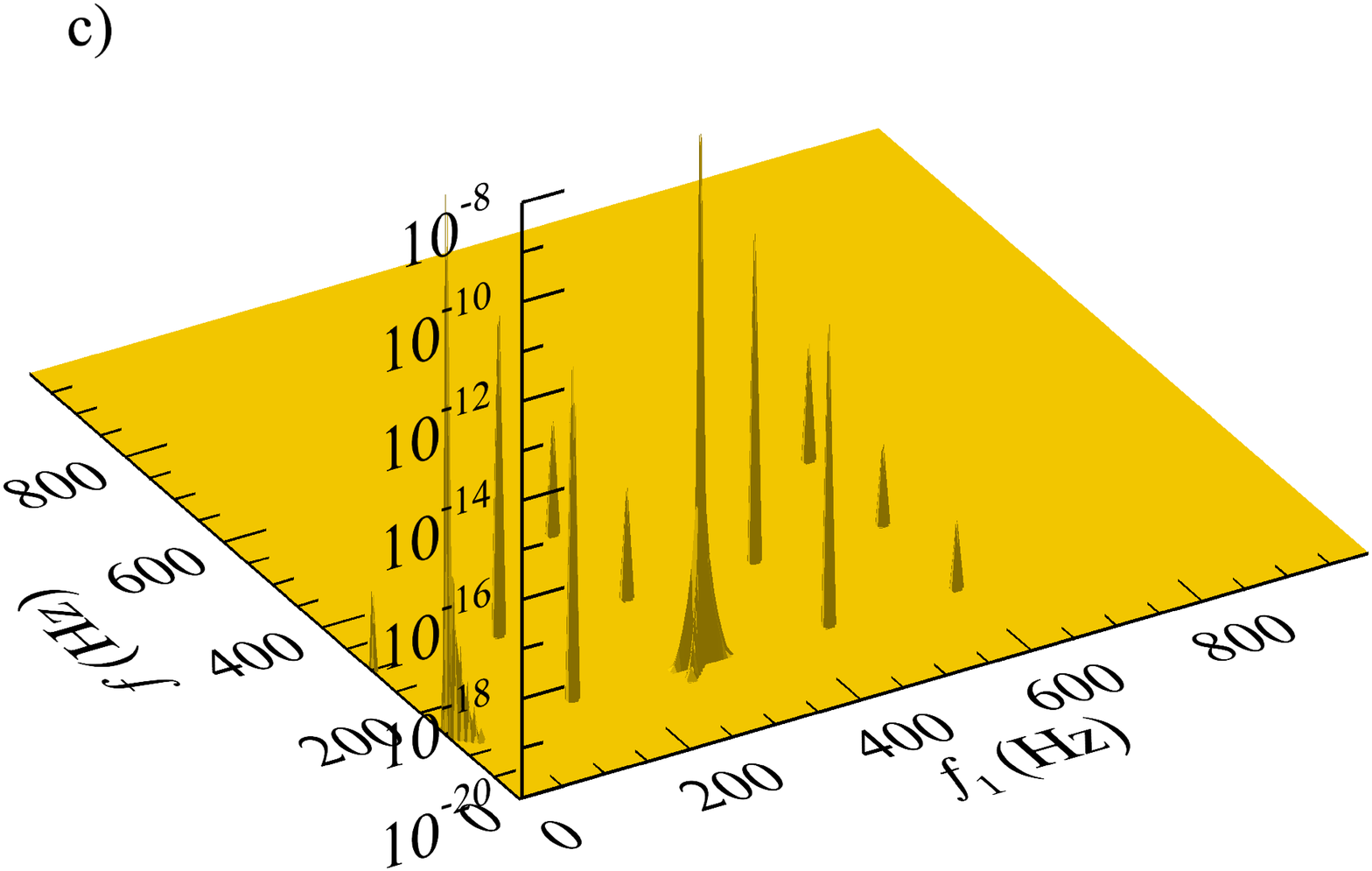}
\end{minipage}%
\begin{minipage}{.5\textwidth}
  \centering
  \includegraphics[width=0.98\linewidth]{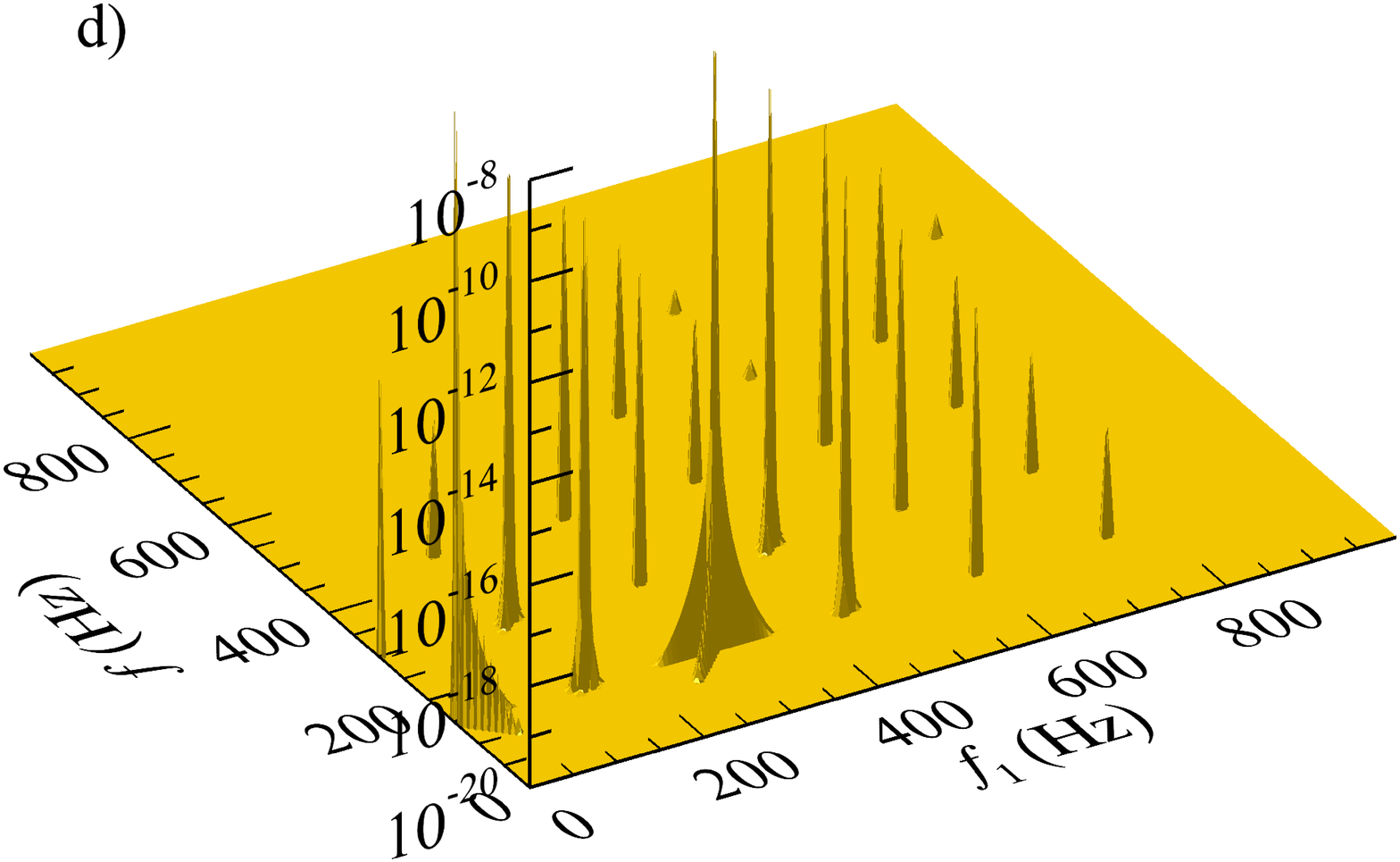}
\end{minipage}
    \caption{Time development of iterative additions to single-frequency triad interactions (first quadrant). (a) 1 iteration, (b) 200 iterations, (c) 300 iterations and (d) 400 iterations.}
    \label{fig:11}
\end{figure*}

Figures~\ref{fig:11}\textit{a-d} show surface plots of the additional contributions (i.e. relative increases) to the power spectrum after 1, 200, 300 and 400 iterations, respectively. These plots show the relative heights of the interactions on an arbitrary logarithmic scale. The resulting addition to the power spectrum is found by summing the contributions along a given temporal frequency, $f$. The corresponding spatial frequency is given by $k=2\pi f$. \clara{}{Only one quadrant is presented, but equal spectral plots will be present in the other three quadrants.}

\begin{figure}[!h]
    \includegraphics[width=0.85 \linewidth]{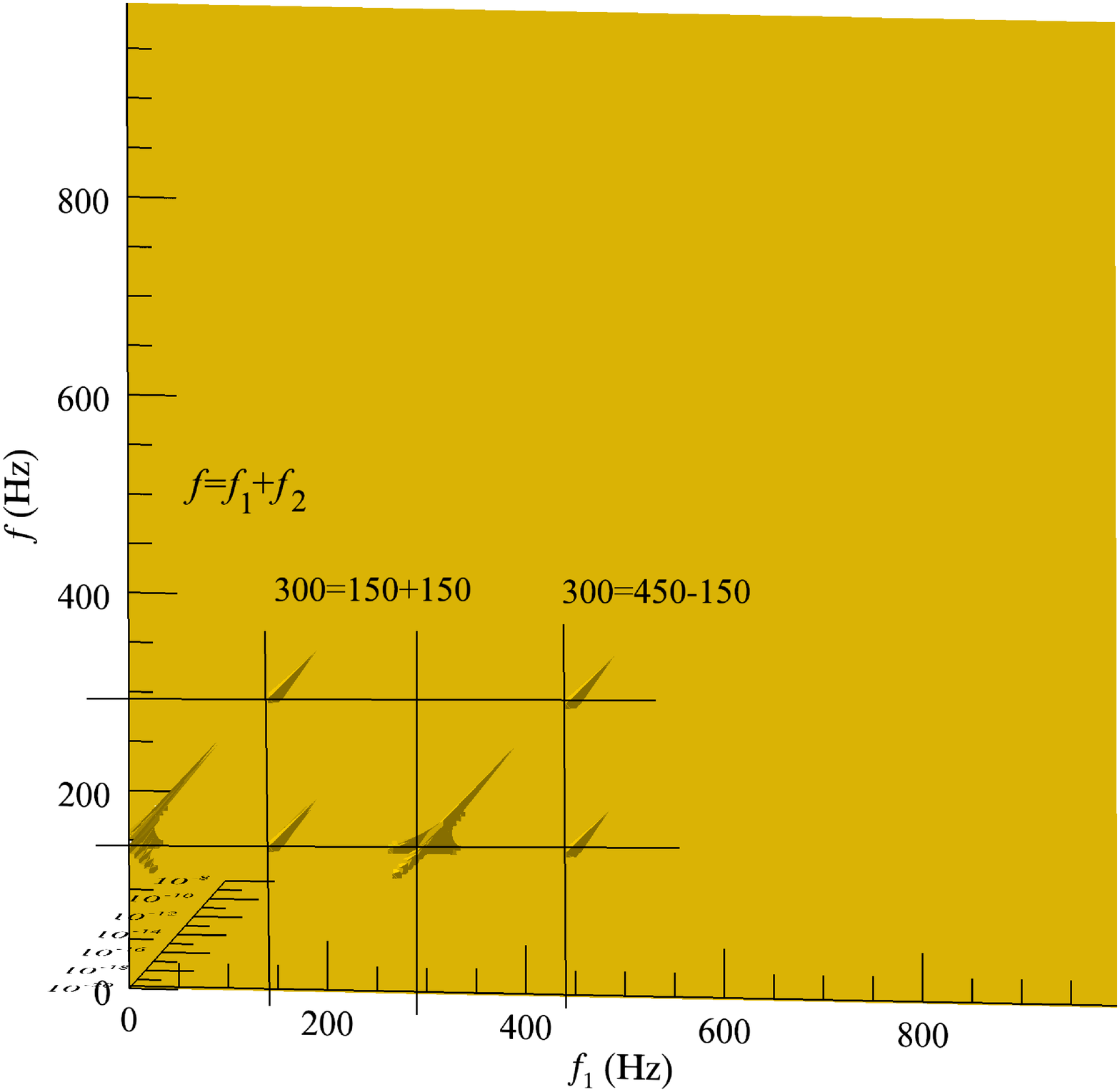}\\
    \caption{Annotated triad combinations at 200 iterations, corresponding to Figure~\ref{fig:11}\textit{b}.}
    \label{fig:12}
\end{figure}

In Figure~\ref{fig:12}, we have shown annotated the frequencies taking part in the generation of the frequency components of the resulting spectrum corresponding to Figure~\ref{fig:11}\textit{b}. The power spectrum addition observed in a measurement is the sum of all contributions contributing to the frequency component labeled with $f$, i.e. the sum over all $f_1$ and $f_2$ frequencies for each selected value of $f$.

\subsection{Dual Rod Results}

\subsubsection{Experiments}\label{sec:dualresults}

\begin{figure}[!h]
    \includegraphics[width=0.85 \linewidth]{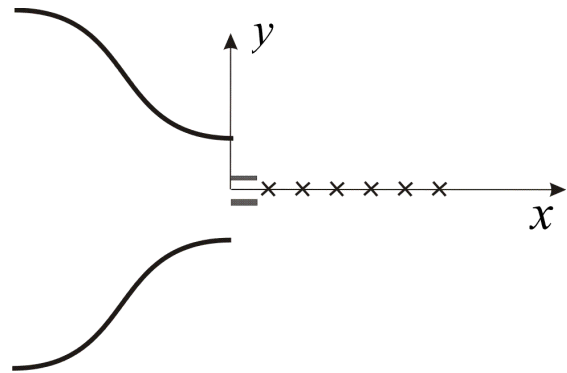}\\
    \caption{Top view of the dual-rod geometry. Rod cross sections: Rod 1 $8 \times 2.5\, mm$ and rod 2 $8 \times 2.8 \, mm$. Measurement positions: $y=0\, mm$ and $x=10$, $20$, $30$, $40$, $50$ and $60 \, mm$.} 
    \label{fig:5}
\end{figure}

\begin{figure}[!h]
    \includegraphics[width=0.85 \linewidth]{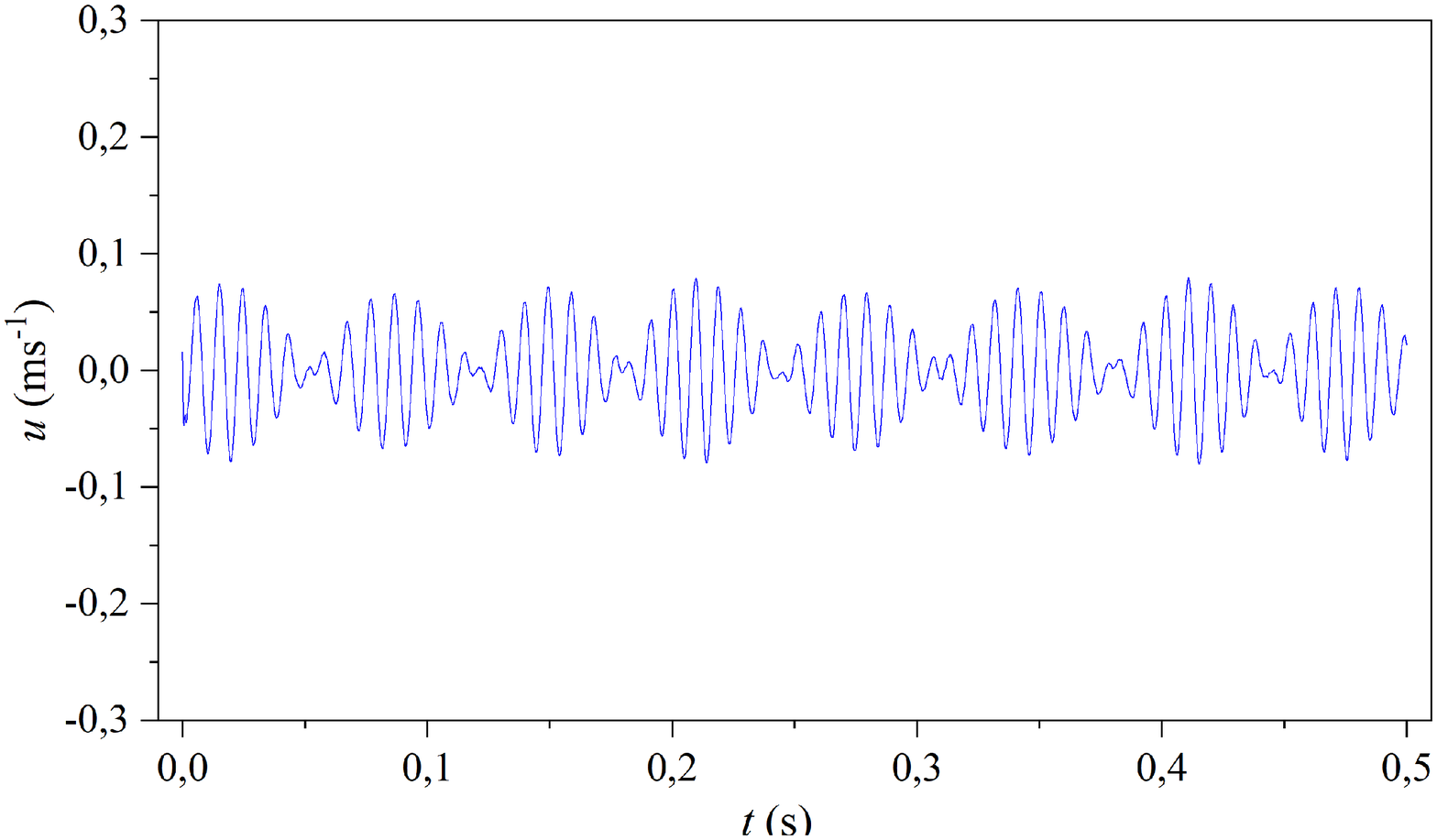}\\
    \caption{A $0.5\,s$ time trace of the velocity fluctuations at $x=10\, mm$, $y=0 \, mm$ and $z=0\, mm$.}
    \label{fig:6}
\end{figure}

The dual-rod geometry is shown in Figure~\ref{fig:5}. The rods had the following cross sections: 
\begin{itemize}
    \item Rod 1: $8 \times 2.5\, mm$
    \item Rod 2: $8 \times 2.8 \, mm$.
\end{itemize} 
The rods were placed symmetrically around the centerline with their respective centers at $y= \pm 2.5\, mm$. \clara{}{This was from empirical investigation found to be the optimal distance between the rods, creating two separate vortex streets near the rods, yet allowing the vortices to start interacting not too far downstream.} Measurement positions along the symmetry axis ($y=0 \,mm$ and $z=0\, mm$) were found optimal in the positions $x=10$, $20$, $30$, $40$, $50$ and $60\, mm$. The fluctuating (AC) part of the velocity signal is shown in Figure~\ref{fig:6}. The beating between the two vortex shedding frequencies, $f_1=99\, Hz$ and $f_2=115\, Hz$, is clearly evident from the velocity time trace. 

\begin{figure*}
\centering
\begin{minipage}{0.5\textwidth}
  \centering
  \includegraphics[width=0.99\linewidth]{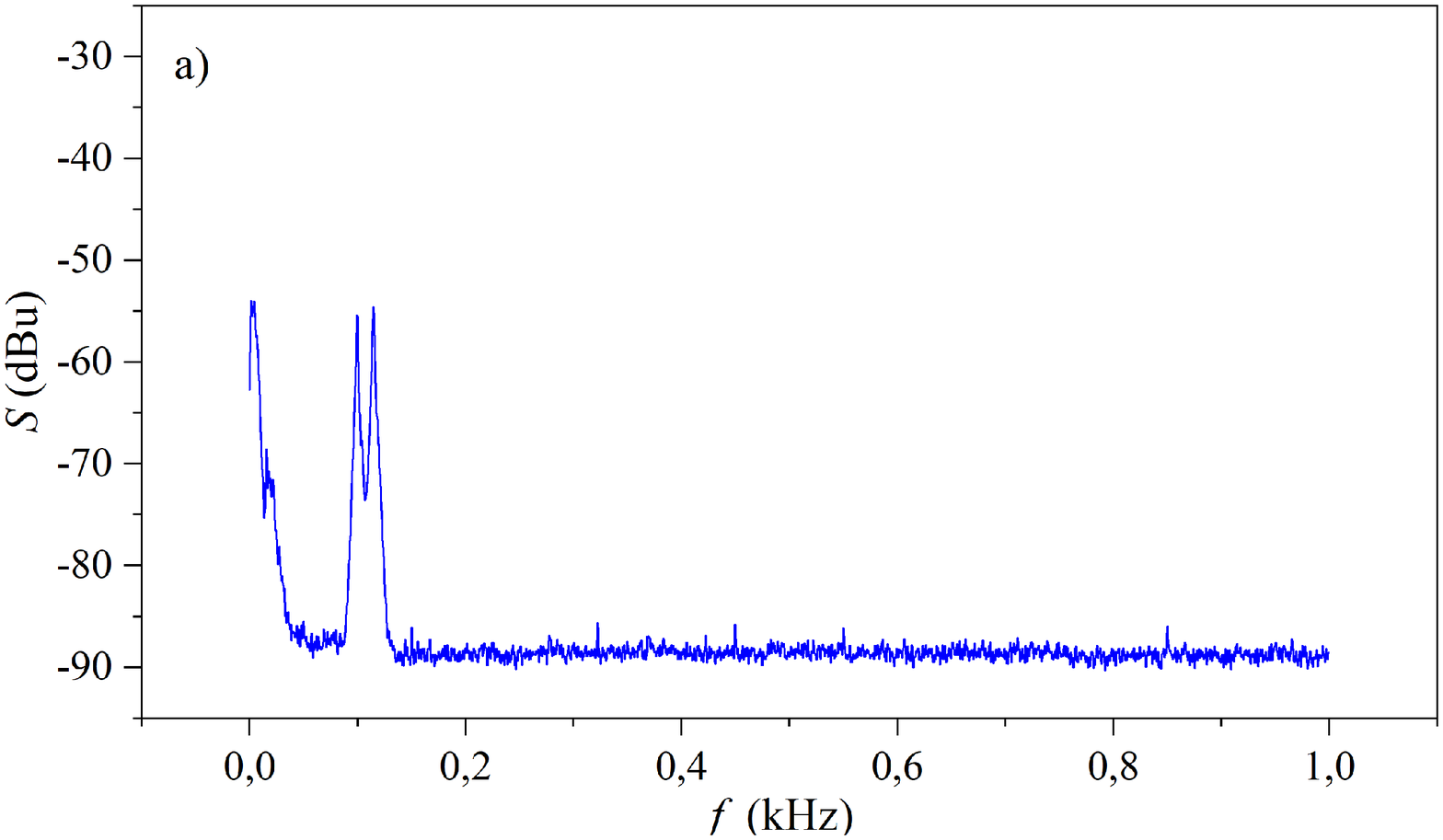}
\end{minipage}%
\begin{minipage}{.5\textwidth}
  \centering
  \includegraphics[width=0.98\linewidth]{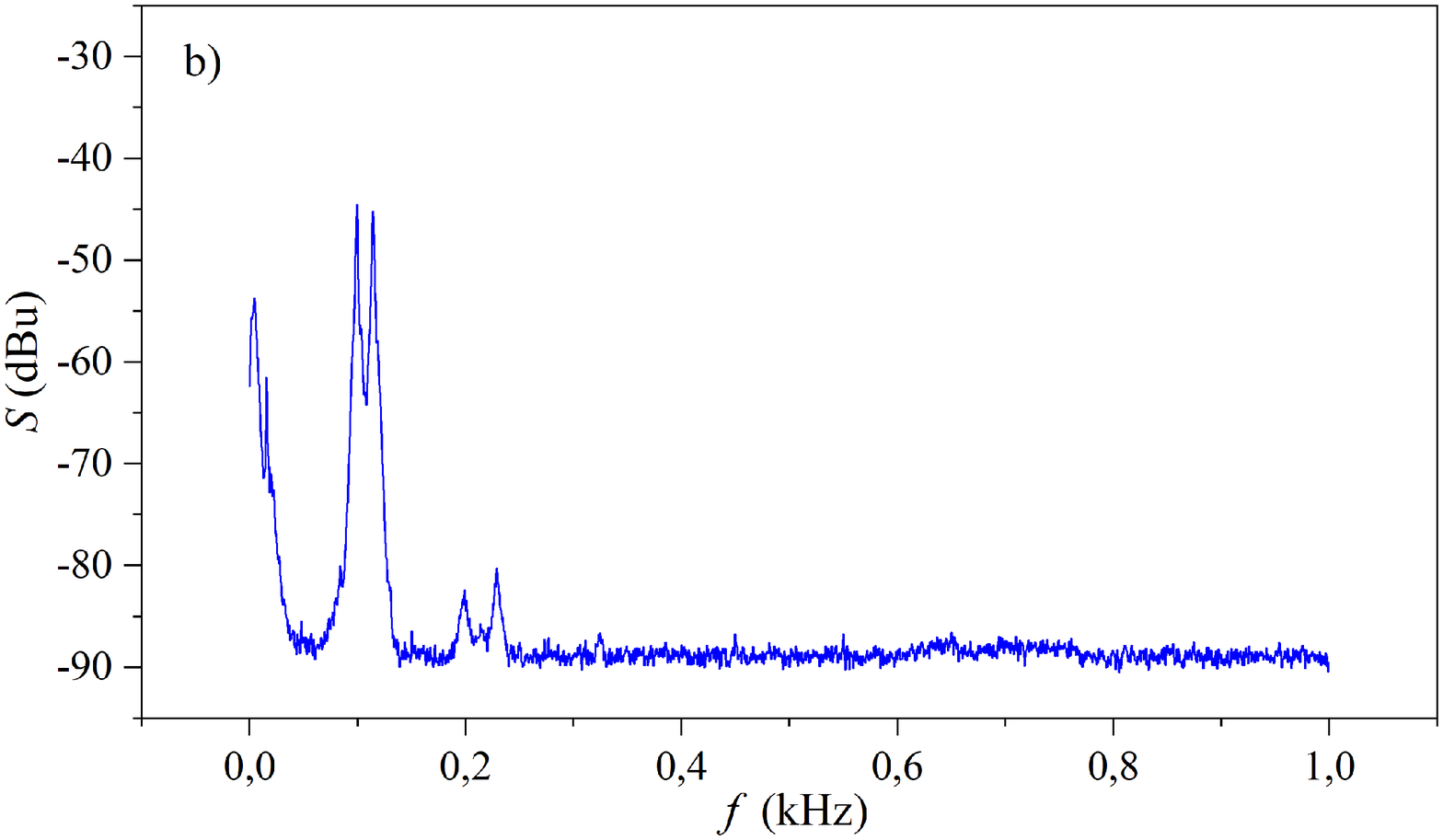}
\end{minipage}
\begin{minipage}{0.5\textwidth}
  \centering
  \includegraphics[width=0.99\linewidth]{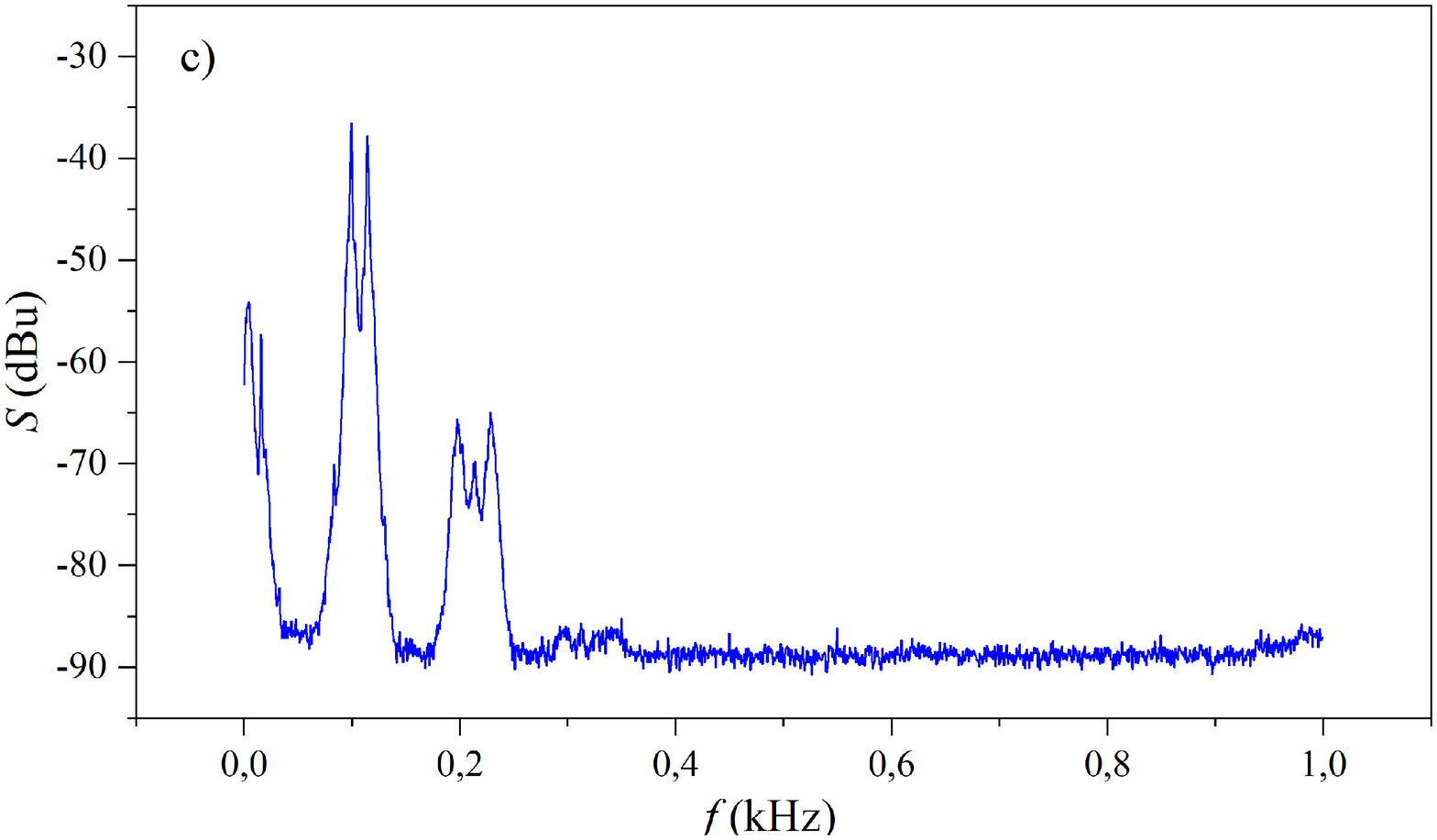}
\end{minipage}%
\begin{minipage}{.5\textwidth}
  \centering
  \includegraphics[width=0.98\linewidth]{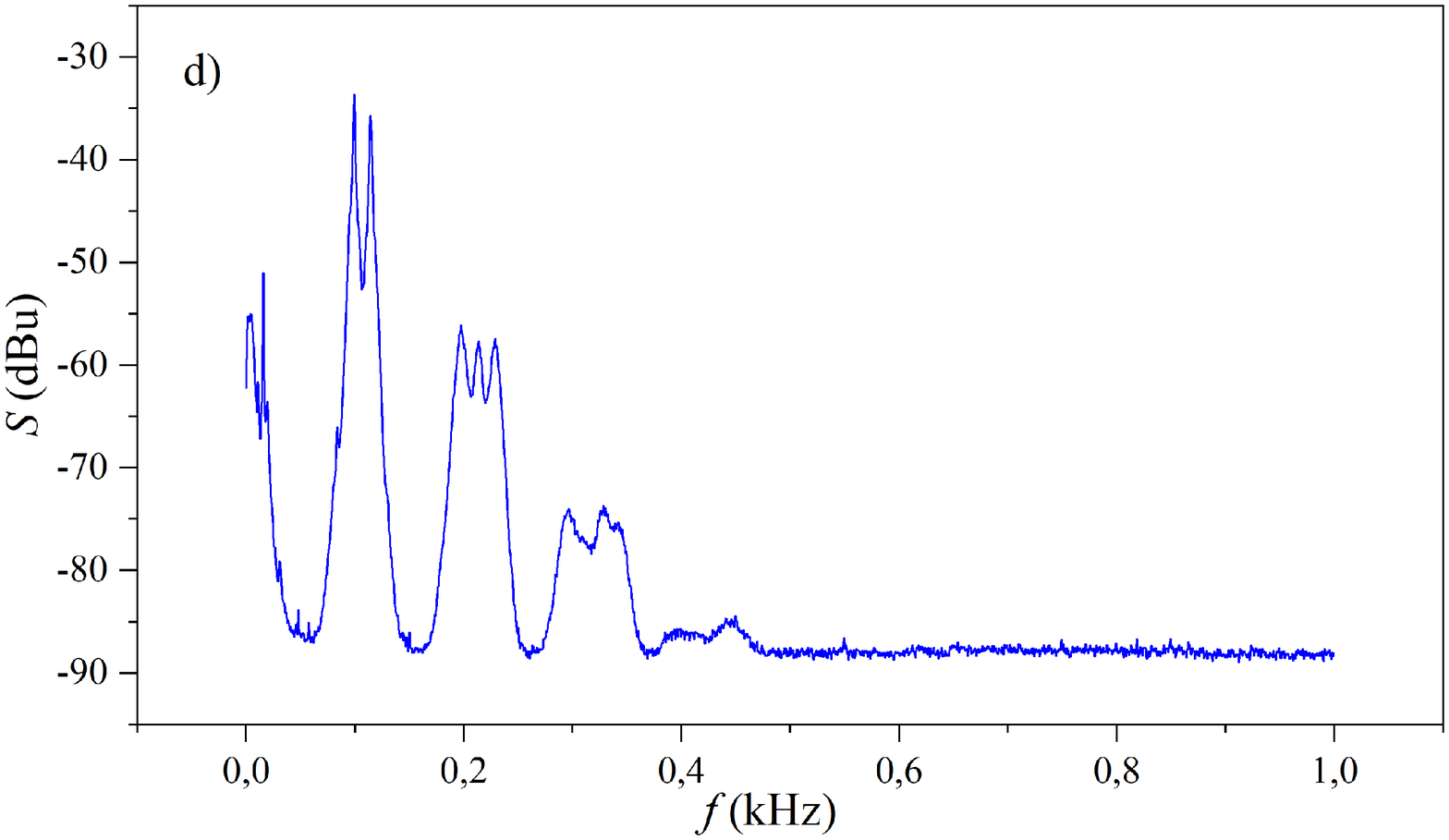}
\end{minipage}
\begin{minipage}{0.5\textwidth}
  \centering
  \includegraphics[width=0.99\linewidth]{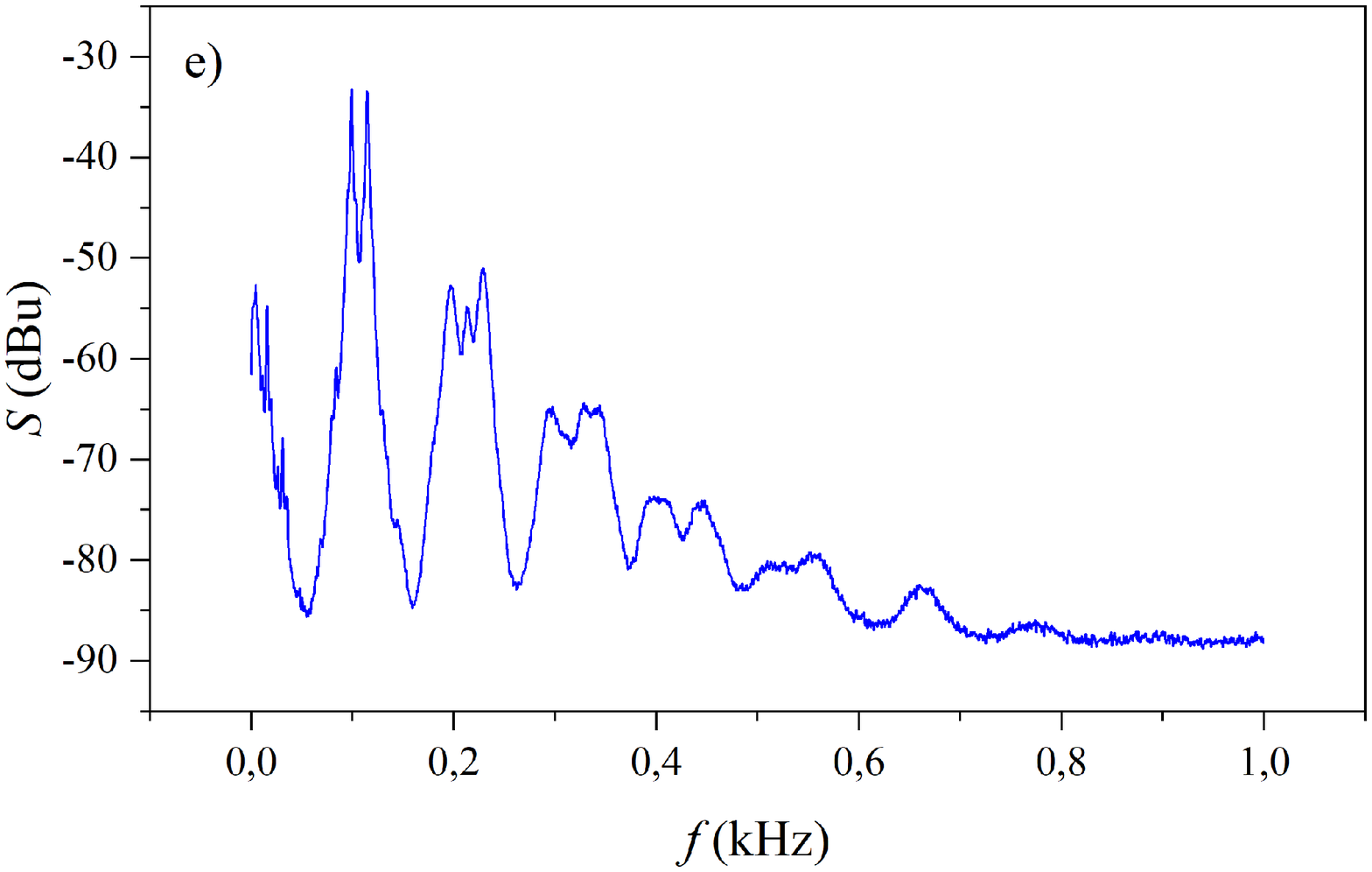}
\end{minipage}%
\begin{minipage}{.5\textwidth}
  \centering
  \includegraphics[width=0.98\linewidth]{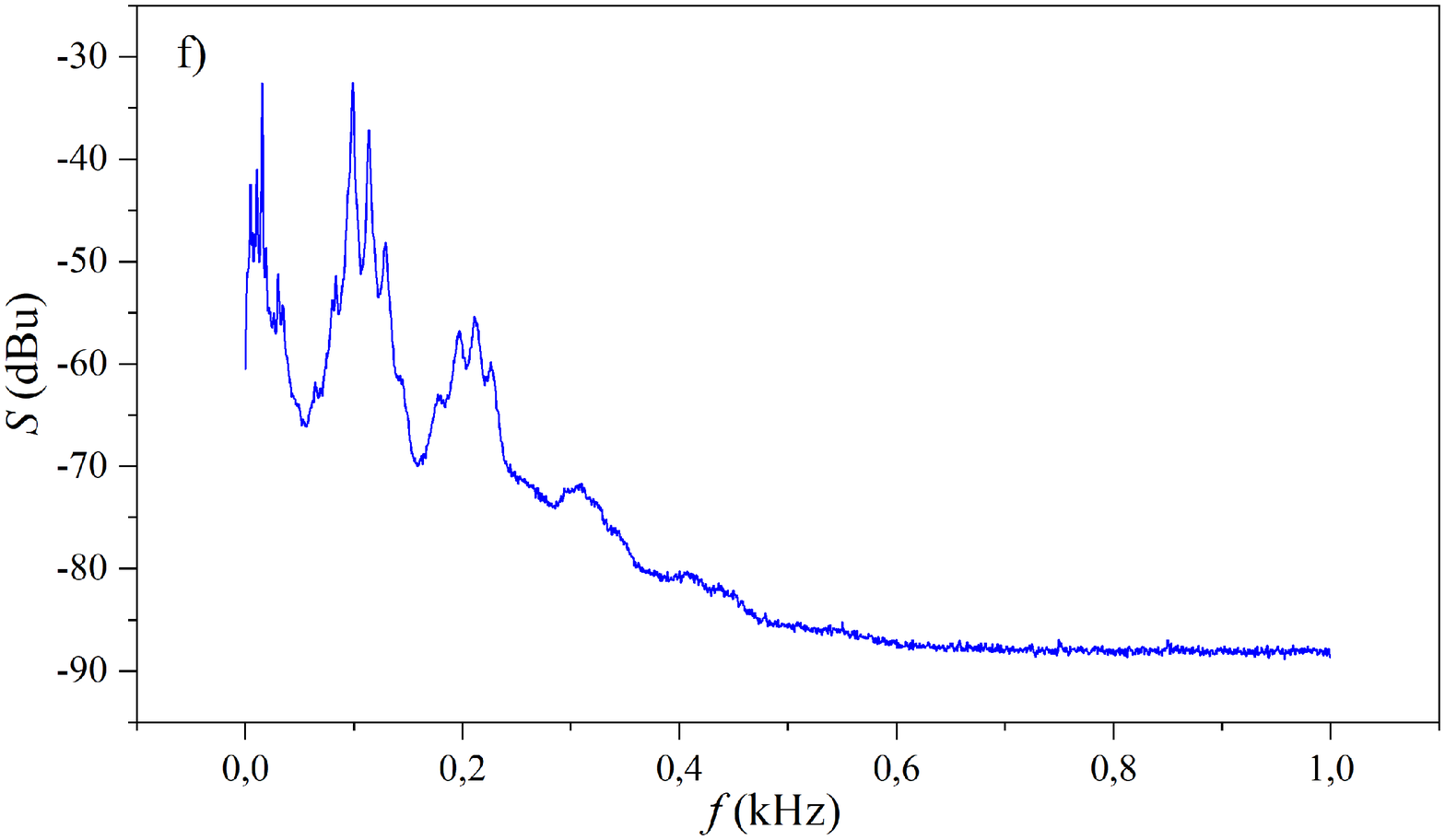}
\end{minipage}
    \caption{Time development of measured dual-rod power spectrum of the hotwire voltage at $y = 0\, mm$ and, respectively, at (a) $x= 10 \, mm$, (b) $x= 20 \, mm$,  (c) $x= 30 \, mm$,  (d) $x= 40 \, mm$,  (e) $x= 50 \, mm$ and  (f) $x= 60 \, mm$.}
    \label{fig:7}
\end{figure*}

Figures~\ref{fig:7}\textit{a–f} show the corresponding power spectra, $S$ with units $(dBu)$, of the voltage signal of the hotwire, measured with the Picoscope. Figure~\ref{fig:7}\textit{a} shows the two distinct vortex frequencies at a position close to the downstream side of the rods at $x=10 \,mm$. \clara{}{Also observed is the expected difference frequency around $f=0$\,Hz.} As we move further downstream, we see a triplet emerging at $x=20\,mm$ in Figure~\ref{fig:7}\textit{b}. The three frequencies are $2f_1$, $f_1+f_2$ and $2f_2$. The next peak to appear will contain all possible combinations of sum frequencies between the first and the second peak group. Many will be identical, and it is not possible to distinguish the weight of each combination separately from the \clara{}{measured} power spectrum. The possible combinations and their respective weights will, however, be clear when we consider the more detailed results of the computer simulations in section~\ref{sec:ComputerSimulations}.

As in the case of a single rod, we again see that the power spectrum at increasing distance downstream, as shown in Figures~\ref{fig:7}\textit{c–f}, approaches an ordinary turbulence power spectrum, both as a result of the growing number of peaks in each group and because of interactions with the background turbulence from the merging shear layers.

\subsubsection{Simulations} \label{sec:ComputerSimulations}

\begin{figure*}
\centering
\begin{minipage}{0.5\textwidth}
  \centering
  \includegraphics[width=0.99\linewidth]{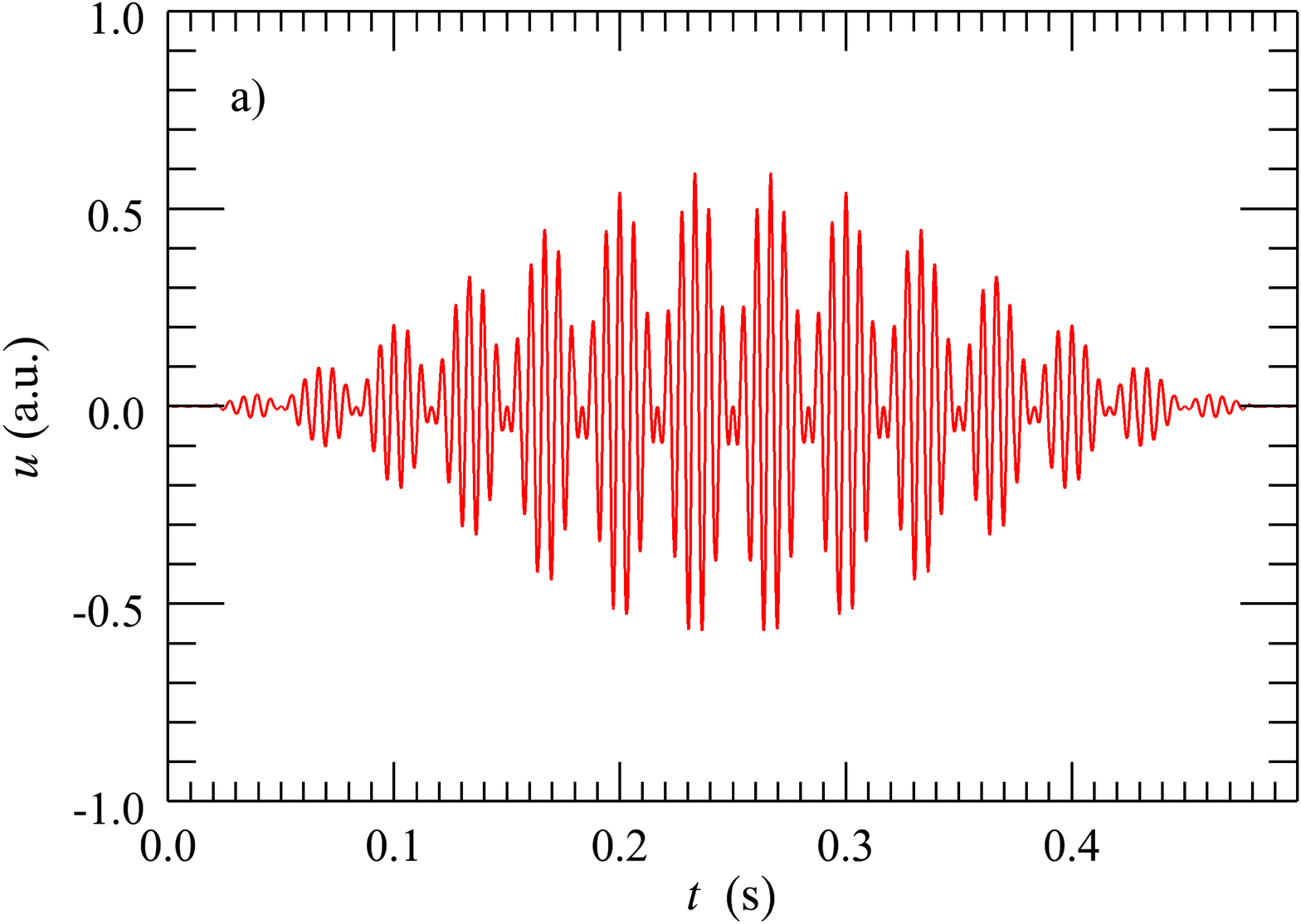}
\end{minipage}%
\begin{minipage}{.5\textwidth}
  \centering
  \includegraphics[width=0.98\linewidth]{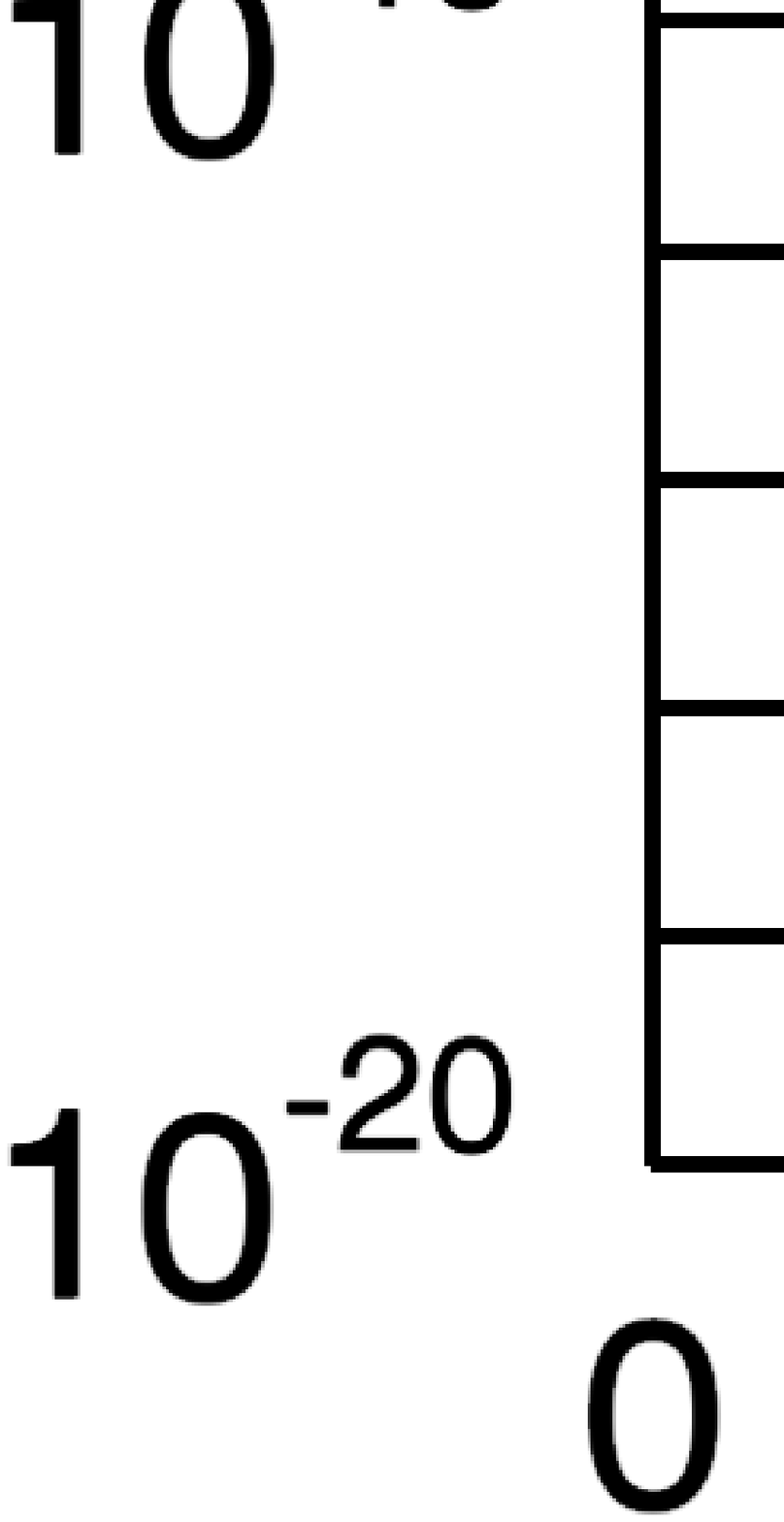}
\end{minipage}
    \caption{(a) A $0.5 \, s$ time record (incident signal) of two Gaussian pulses with modulation frequencies $150\, Hz$ and $180\, Hz$ and (b) the corresponding power spectrum.}
    \label{fig:13}
\end{figure*}

Figure~\ref{fig:13}\textit{a} shows an example of a $0.5\, s$ time record of a two-frequency signal consisting of the sum of two Gaussian pulses modulated with frequencies $f_1=150\, Hz$ and $f_2=180\, Hz$. Figure~\ref{fig:13}\textit{b} shows the corresponding power spectrum. This signal is meant to represent the vortex shedding frequencies close to the dual rods experiment described in Section~\ref{sec:dualresults}.

\begin{figure*}
\centering
\begin{minipage}{0.5\textwidth}
  \centering
  \includegraphics[width=0.99\linewidth]{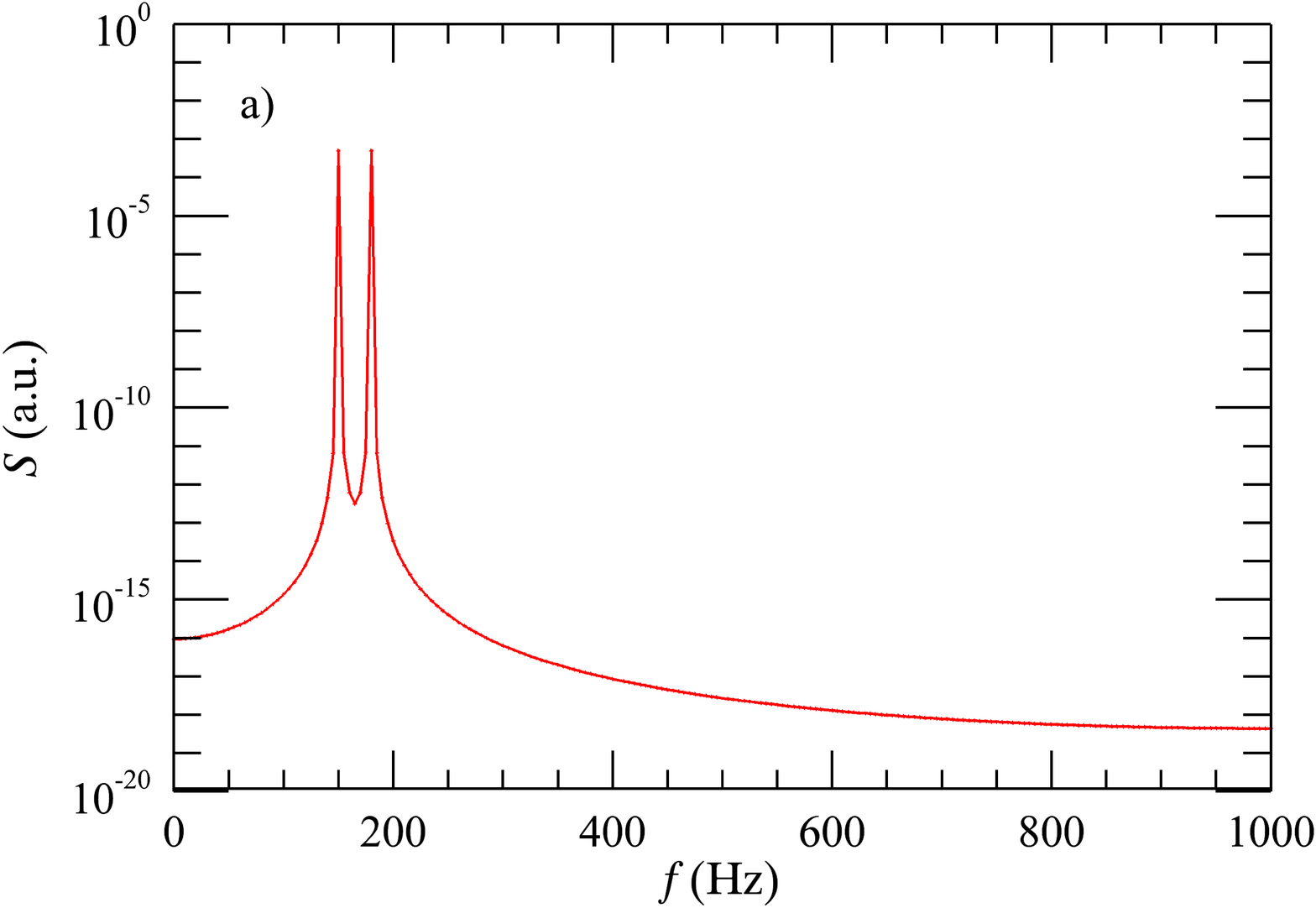}
\end{minipage}%
\begin{minipage}{.5\textwidth}
  \centering
  \includegraphics[width=0.98\linewidth]{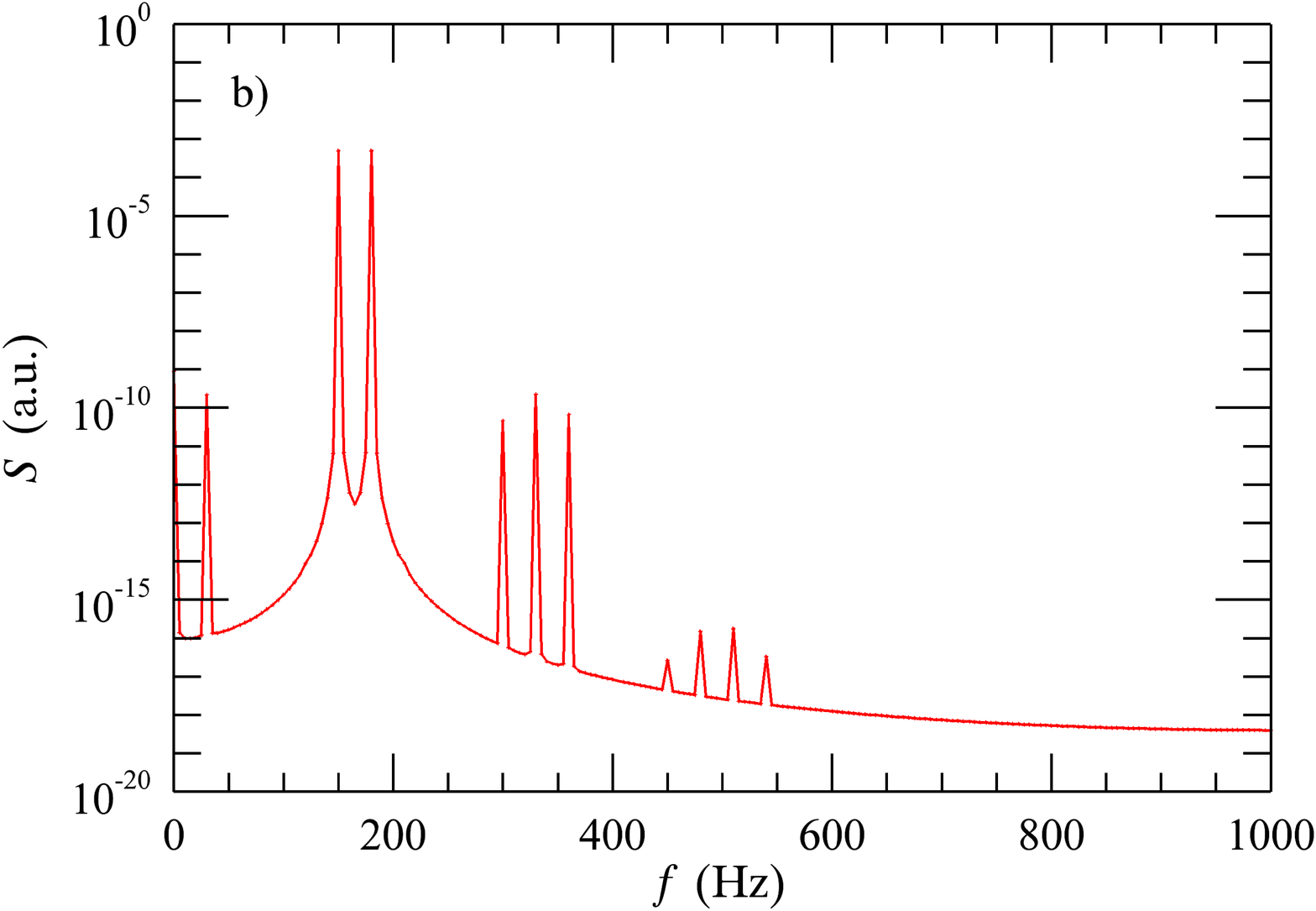}
\end{minipage}
\begin{minipage}{0.5\textwidth}
  \centering
  \includegraphics[width=0.99\linewidth]{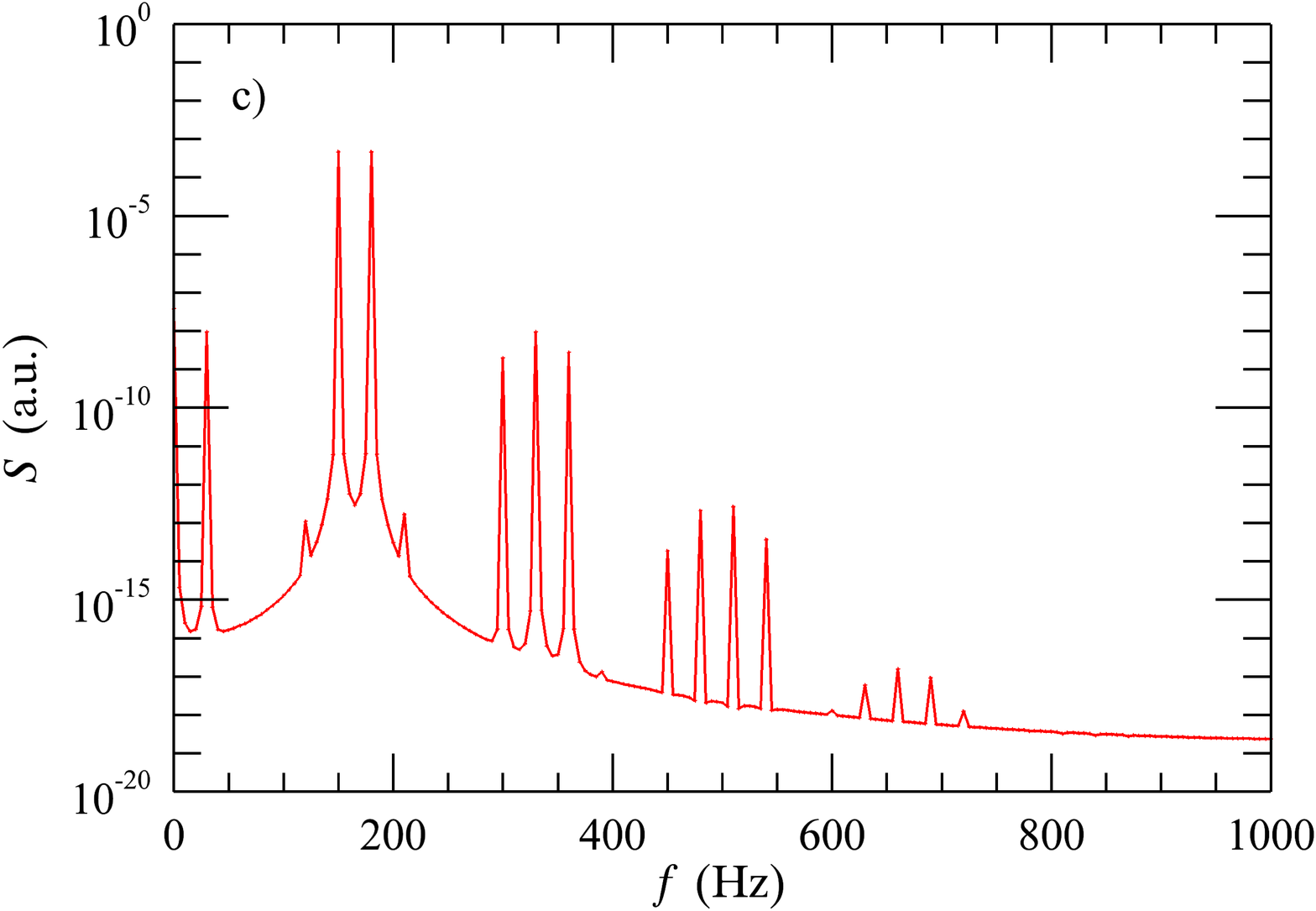}
\end{minipage}%
\begin{minipage}{.5\textwidth}
  \centering
  \includegraphics[width=0.98\linewidth]{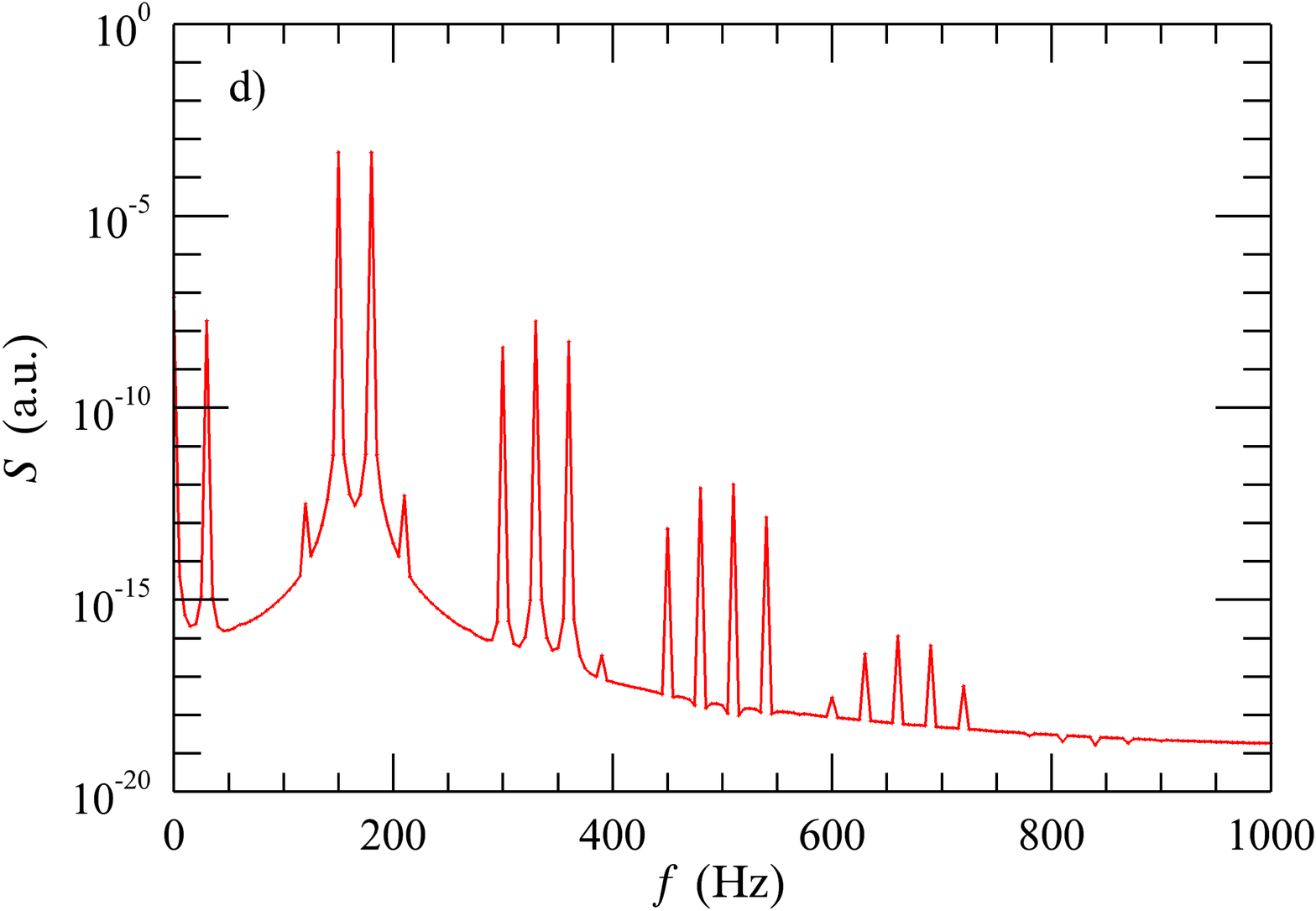}
\end{minipage}
    \caption{Time development of dual-frequency power spectra. (a) 1 iteration, (b) 200 iterations, (c) 300 iterations and (d) 400 iterations.}
    \label{fig:14}
\end{figure*}

Figures~\ref{fig:14}\textit{a-d} show a sequence of power spectra with steps of 100 iterations, a: 1, b: 200, c: 300 and d: 400 iterations. It is clear that each group of frequencies develop more and more peaks with increasing number of iterations, as a result of interactions between their difference frequencies. The frequency difference is also seen at $180\, Hz -150\, Hz =30\, Hz$.

\begin{figure*}
\centering
\begin{minipage}{0.5\textwidth}
  \centering
  \includegraphics[width=0.99\linewidth]{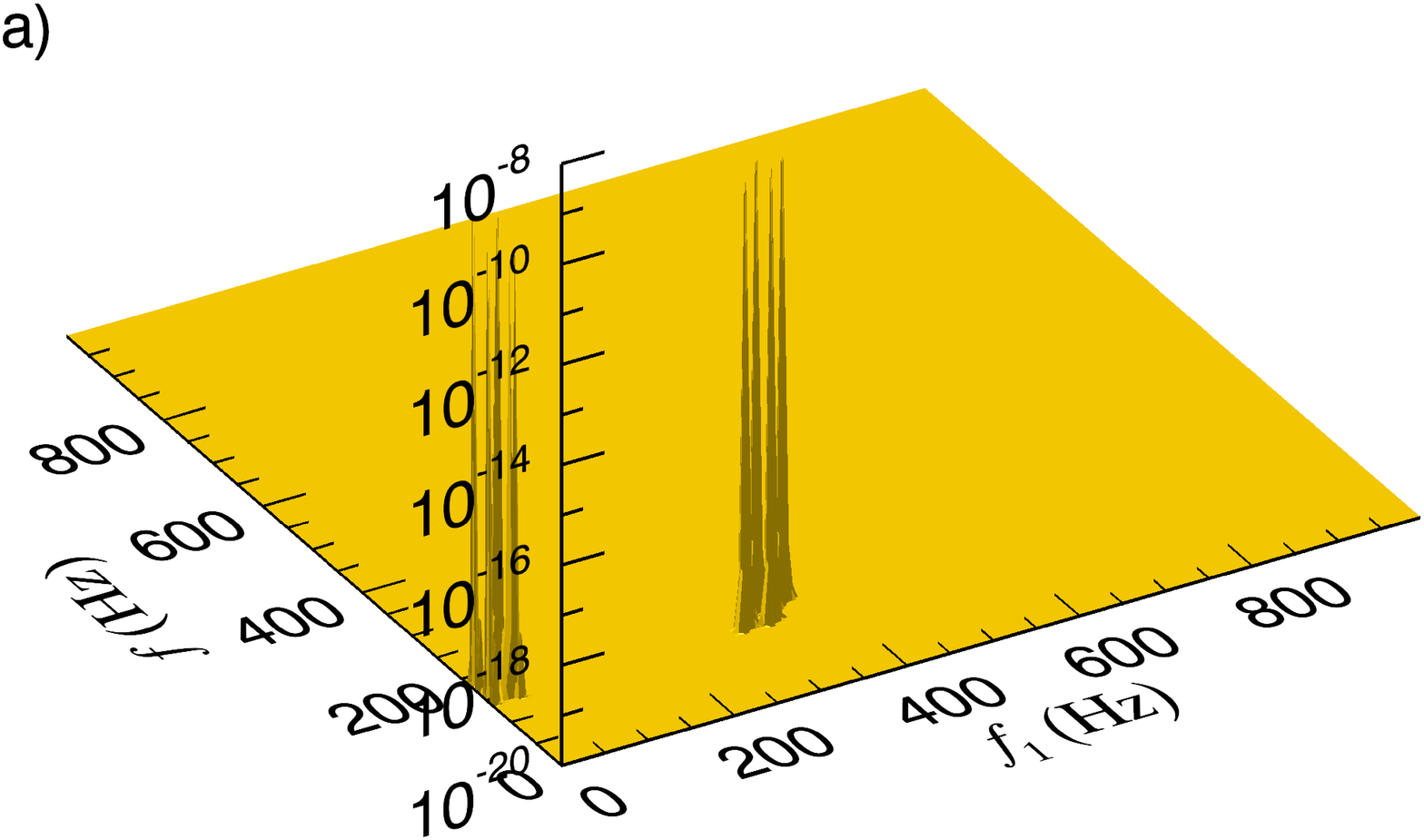}
\end{minipage}%
\begin{minipage}{.5\textwidth}
  \centering
  \includegraphics[width=0.98\linewidth]{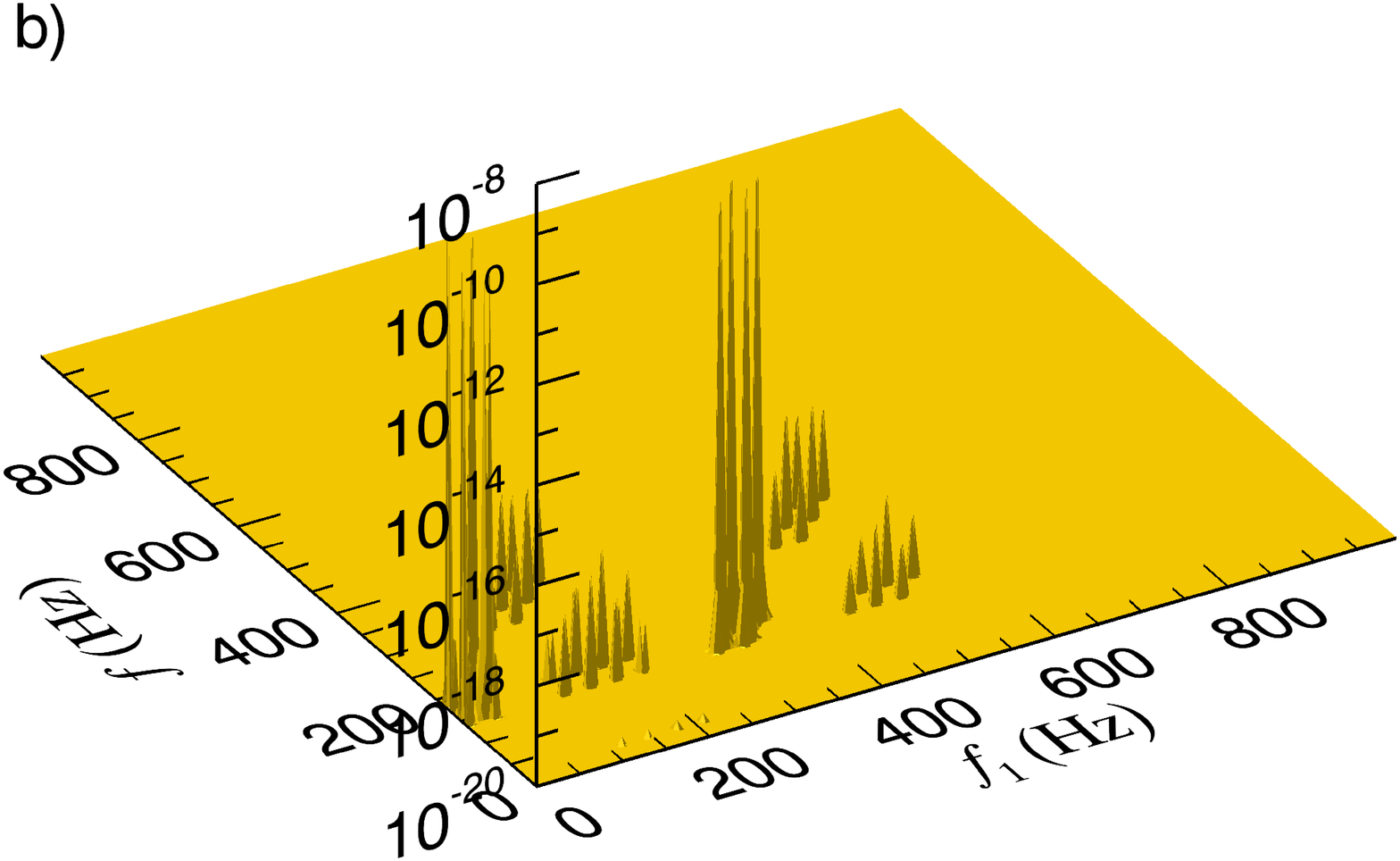}
\end{minipage}
\begin{minipage}{0.5\textwidth}
  \centering
  \includegraphics[width=0.99\linewidth]{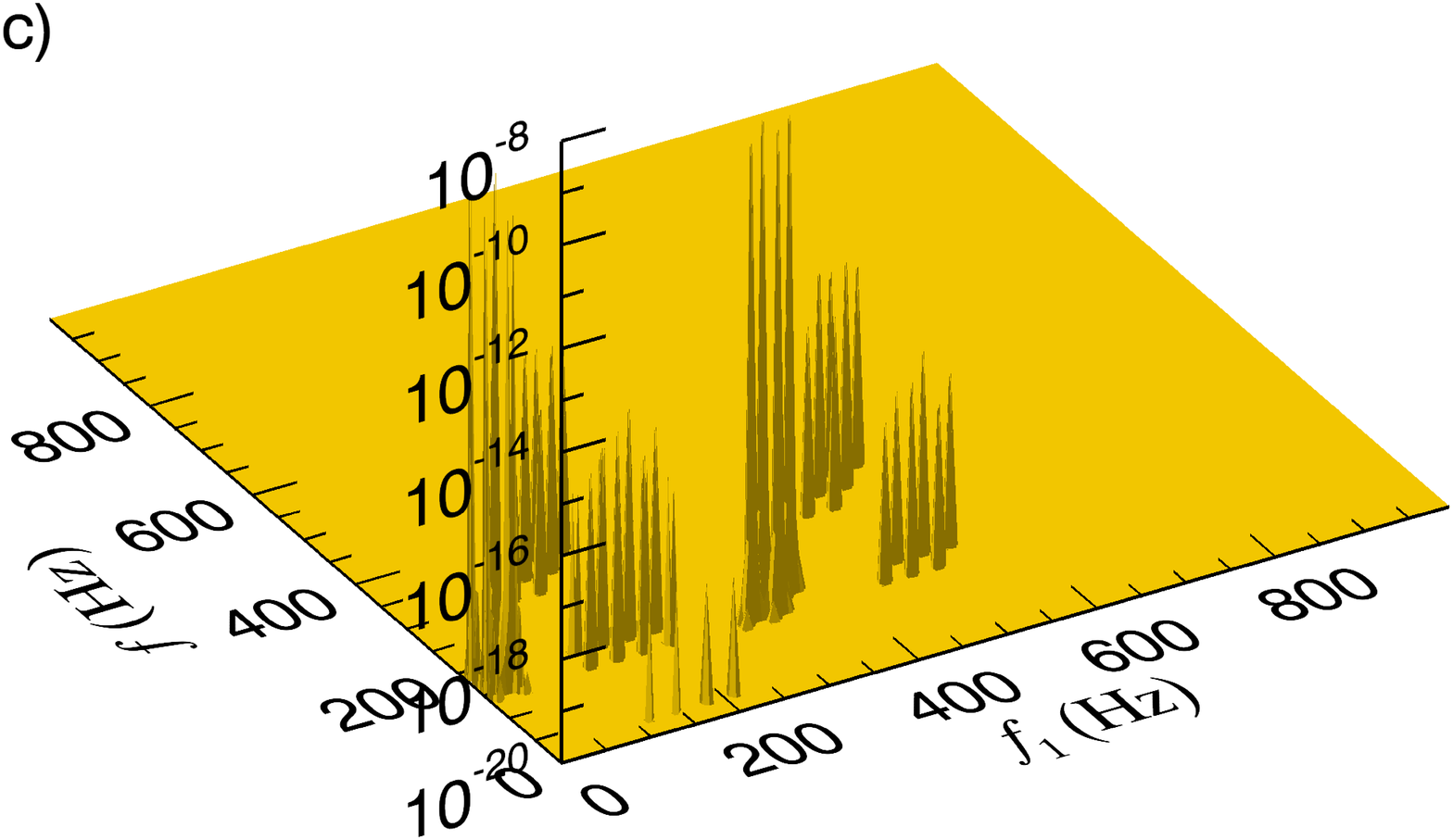}
\end{minipage}%
\begin{minipage}{.5\textwidth}
  \centering
  \includegraphics[width=0.98\linewidth]{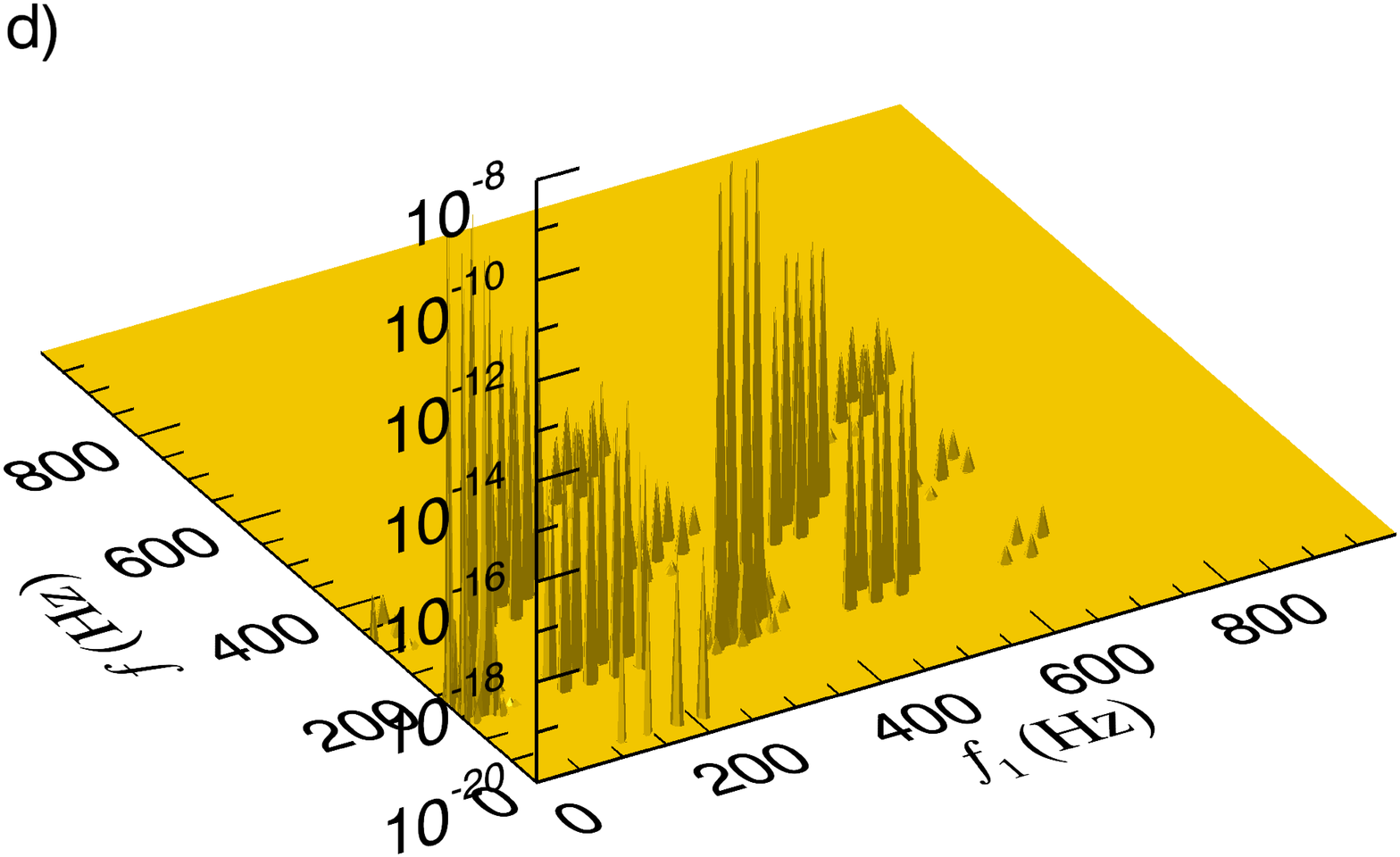}
\end{minipage}
    \caption{Time development of iterative additions to dual-rod triad interactions (first quadrant). (a) 1 iteration, (b) 100 iterations, (c) 200 iterations and (d) 400 iterations.}
    \label{fig:15}
\end{figure*}

Figures~\ref{fig:15}\textit{a–d} show the development of the additional contributions to the power spectrum. \clara{}{The plots show the contribution from a single iteration in the program loop after $p$ iterations.} Again, the total addition to the power spectrum at a given value of the frequency, $f$, is found by summing all contributions yielding the same value of $f$. 

\begin{figure*}
\centering
\begin{minipage}{0.5\textwidth}
  \centering
  \includegraphics[width=0.99\linewidth]{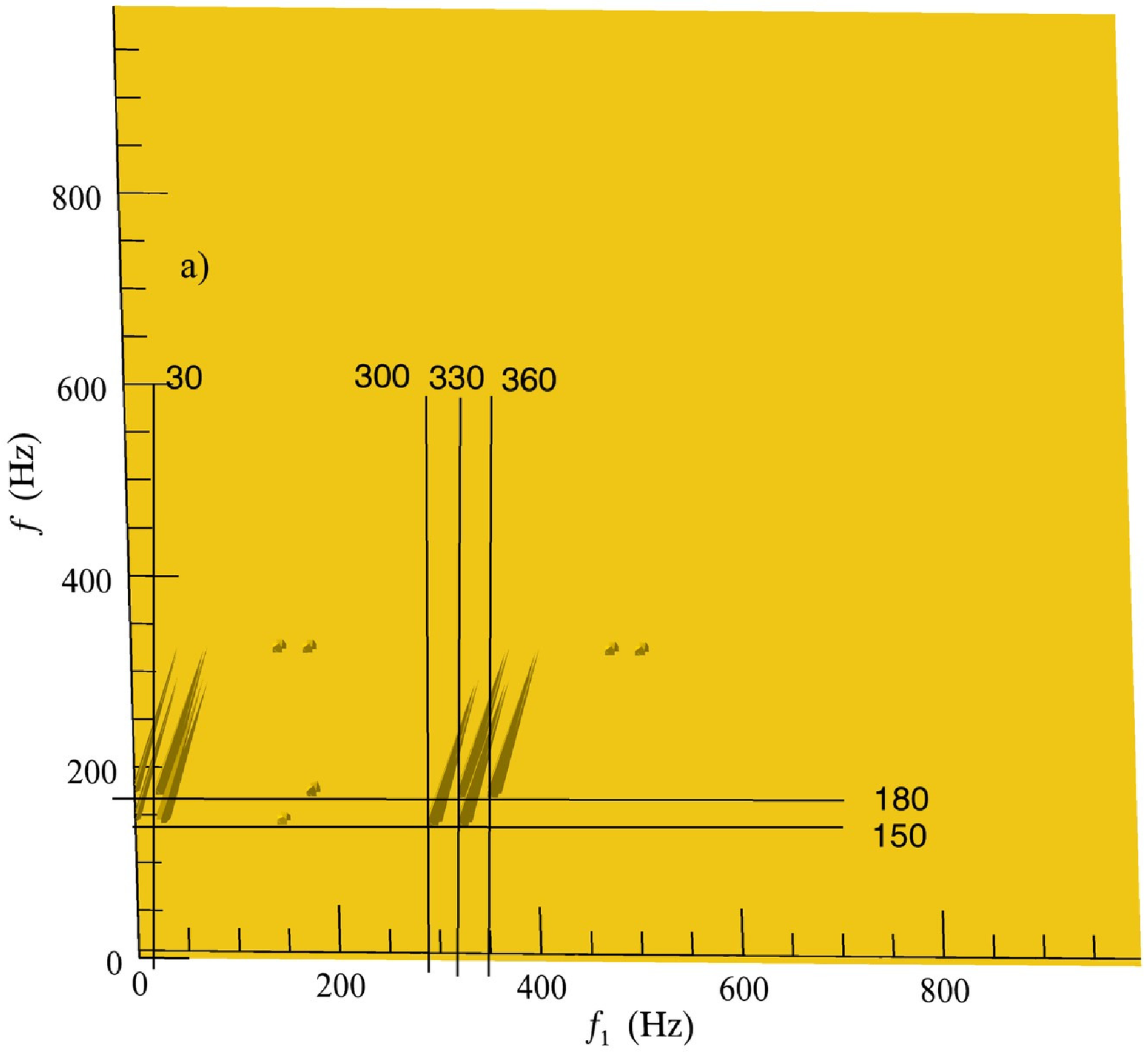}
\end{minipage}%
\begin{minipage}{.5\textwidth}
  \centering
  \includegraphics[width=0.98\linewidth]{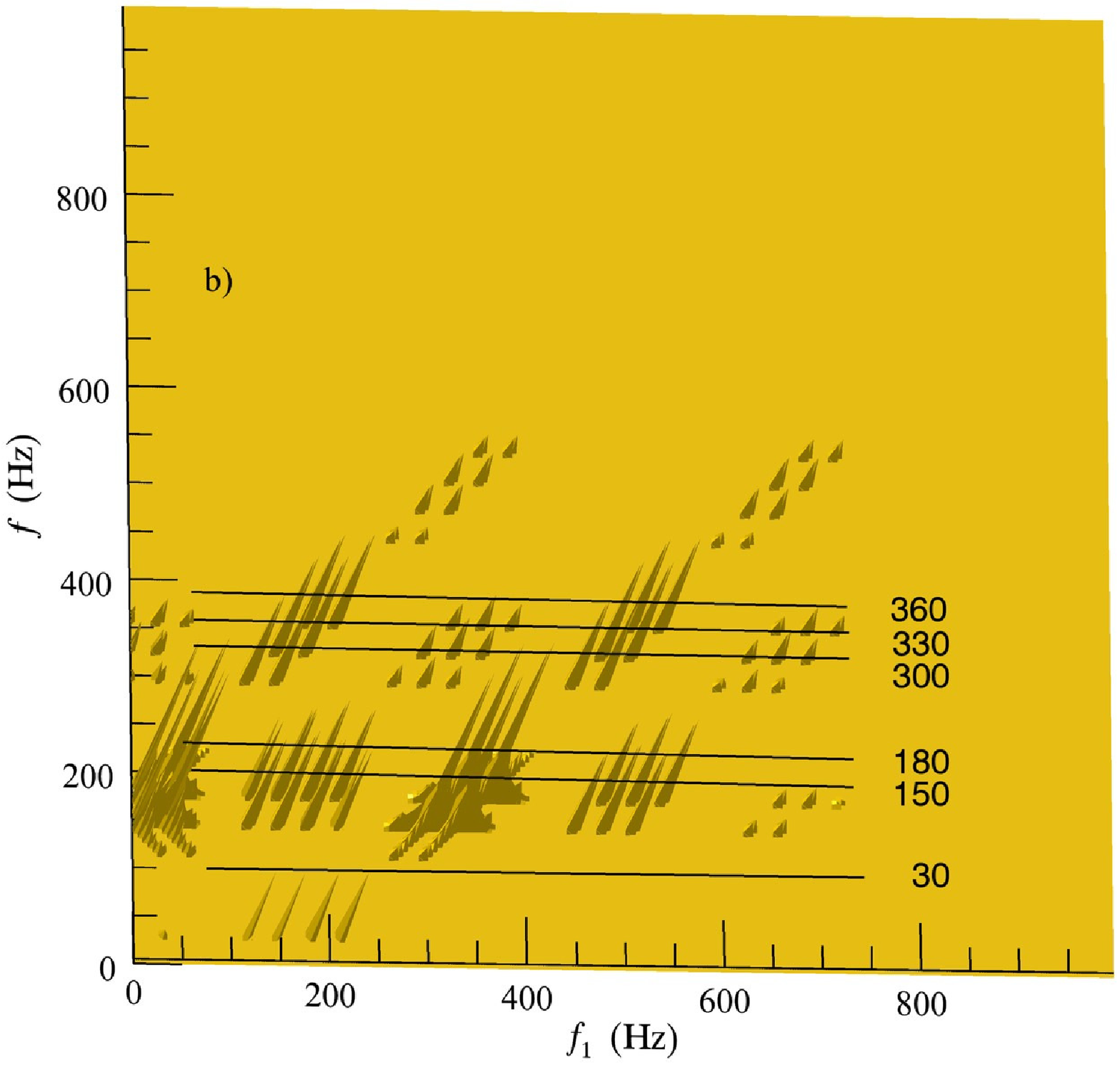}
\end{minipage}
    \caption{Annotated dual-frequency triad combinations. (a) Initial contributions at the incident frequencies. (b) After 400 iterations, showing many possible triad interactions.}
    \label{fig:16}
\end{figure*}

In Figure~\ref{fig:16}, we have marked some of the frequency combinations that contribute to the power spectrum peaks displayed in Figures~\ref{fig:15}\textit{a} and~\ref{fig:15}\textit{d}.

\clara{}{Figures~\ref{fig:14} and Figure~\ref{fig:15} offer valuable insight into the effects of the nonlinear transport term. In Figure~\ref{fig:7}, we see the effect of additional turbulence developing downstream in the jet, especially the increased base level in Figure~\ref{fig:7}f. This background level makes it difficult to compare the levels of the harmonics.}

\clara{}{The reason why the difference peak and the three second harmonic peaks are nearly equal, as we see clearly in Figure~\ref{fig:14}, is that the difference peak is created, not just from the difference between the ``primary modes'', but also from differences in all higher order groups of peaks. This is not surprising, since the differences between the peaks in the different peak groups are the same (which tendency is also clearly seen if one keeps iterating on subsequent interactions than those initiated in eqn.~\ref{eq:HSexample}). Our computer program allows us to see directly the strengths of the various combinations, which modes enter (Figure~\ref{fig:15} and~\ref{fig:16}) and even in what sequence. We consider the comparison of the experimental results with the computer-generated ones presented in Figure~\ref{fig:11} and~\ref{fig:15} the main result of the paper.}

\subsubsection{Simulations with measured velocity signal as input}

Figure~\ref{fig:6} shows one record of the 100 measured dual-rod time records measured close to the jet exit. The signal clearly shows the beats between the two main vortex shedding frequencies and also spurious fluctuations and the AC-value of the signal is only a few percent of the total.

\begin{figure*}
\centering
\begin{minipage}{0.5\textwidth}
  \centering
  \includegraphics[width=0.99\linewidth]{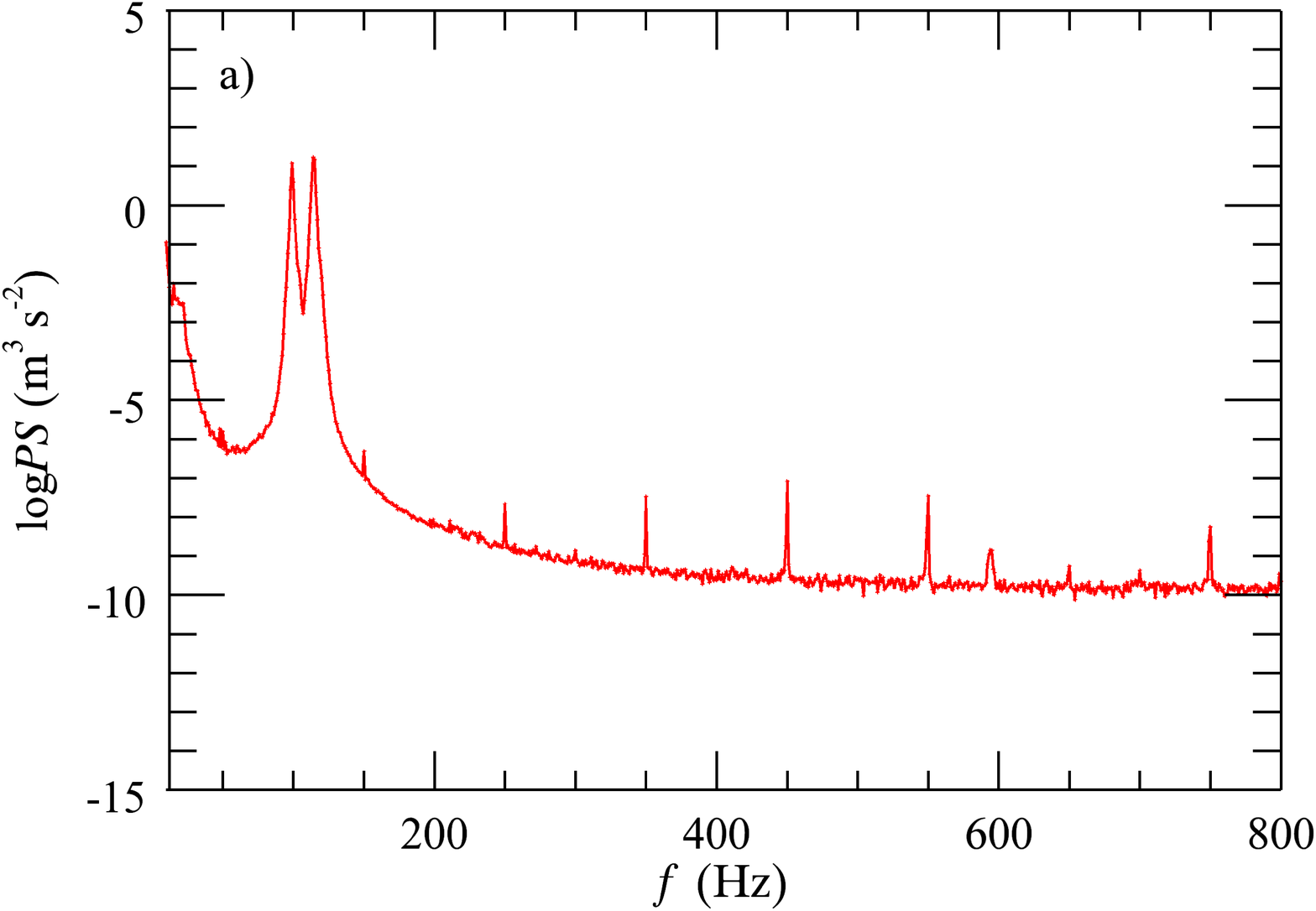}
\end{minipage}%
\begin{minipage}{.5\textwidth}
  \centering
  \includegraphics[width=0.98\linewidth]{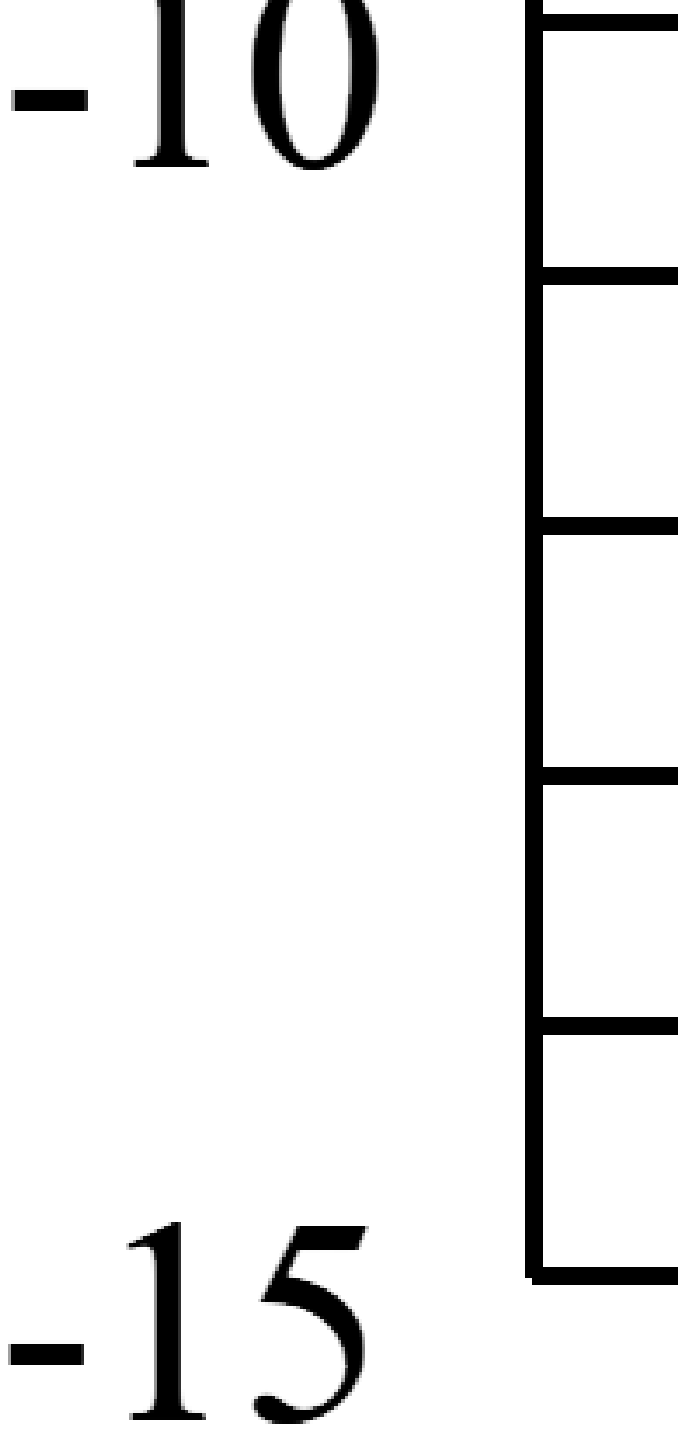}
\end{minipage}
\begin{minipage}{0.5\textwidth}
  \centering
  \includegraphics[width=0.99\linewidth]{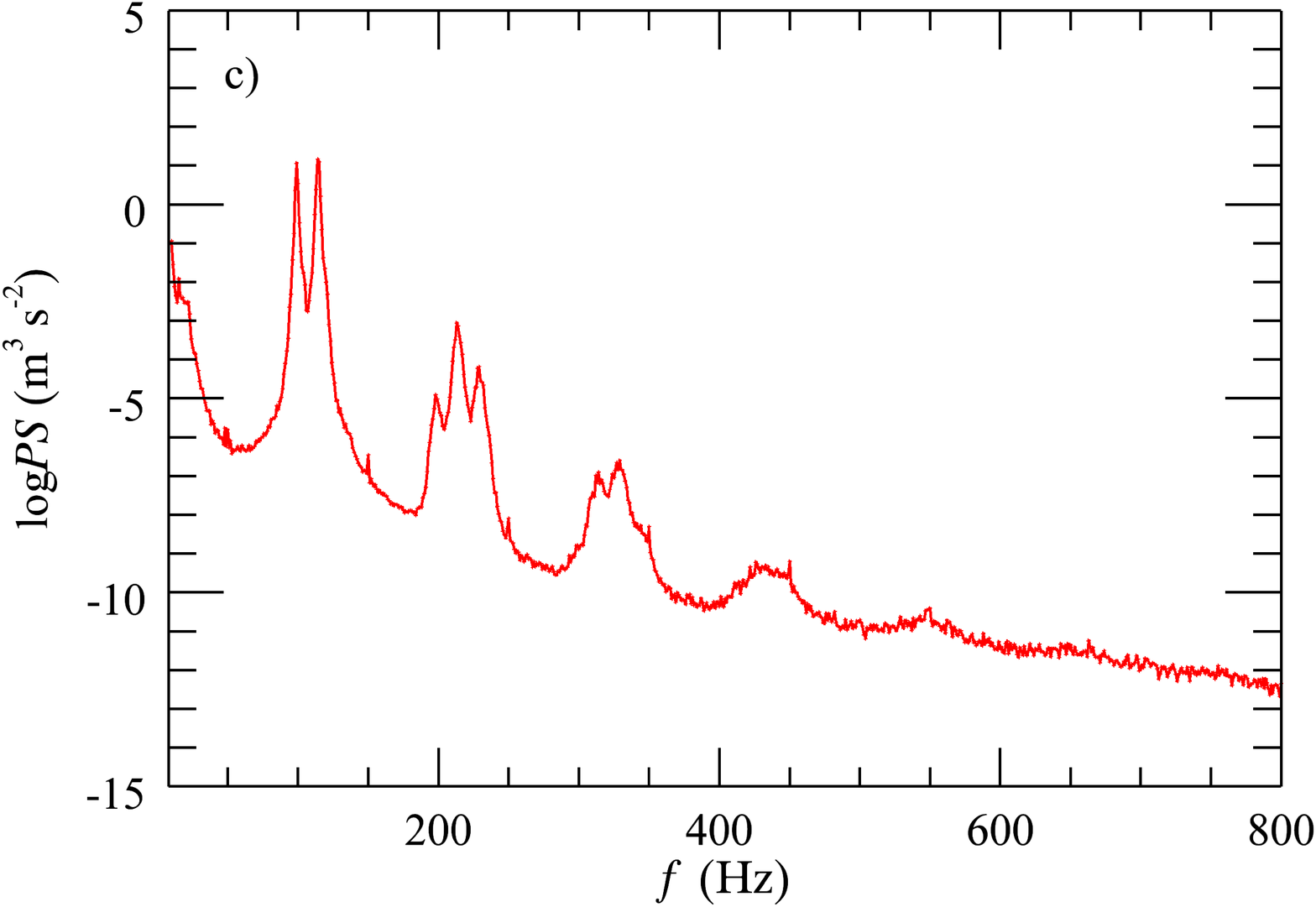}
\end{minipage}%
\begin{minipage}{.5\textwidth}
  \centering
  \includegraphics[width=0.98\linewidth]{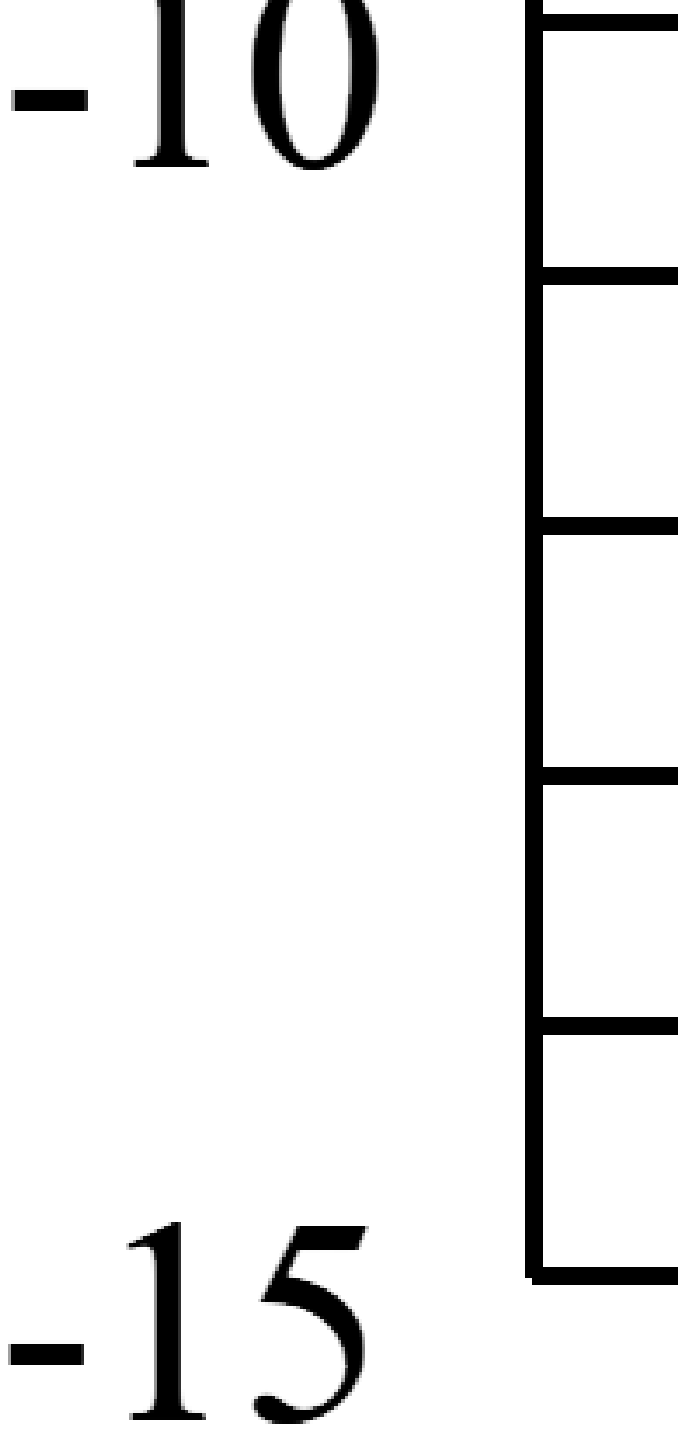}
\end{minipage}
    \caption{Time development of measured dual-rod power spectra. (a) 10 iterations, (b) 20 iterations, (c) 30 iterations and (d) 40 iterations.}
    \label{fig:18}
\end{figure*}

Figure~\ref{fig:18} shows a sequence of power spectra computed with the Navier-Stokes program using this signal measured near the rods as input and averaged over 100 records. The figure shows the development of the group of peaks as in Figure~\ref{fig:14}, but the detailed structure is obscured by unsteadiness in the generated vortex frequencies in the real flow. However, the plots illustrate quite clearly both the sum and difference frequency peaks predicted (including the expected $f=0$\,Hz peak), as well as the net transfer of kinetic energy to higher frequencies. Even within each group of frequencies, it can be seen how interactions are biased to the high frequency side by the special form of the nonlinearity in the Navier-Stokes equation. 

\begin{figure*}
\centering
\begin{minipage}{0.5\textwidth}
  \centering
  \includegraphics[width=0.99\linewidth]{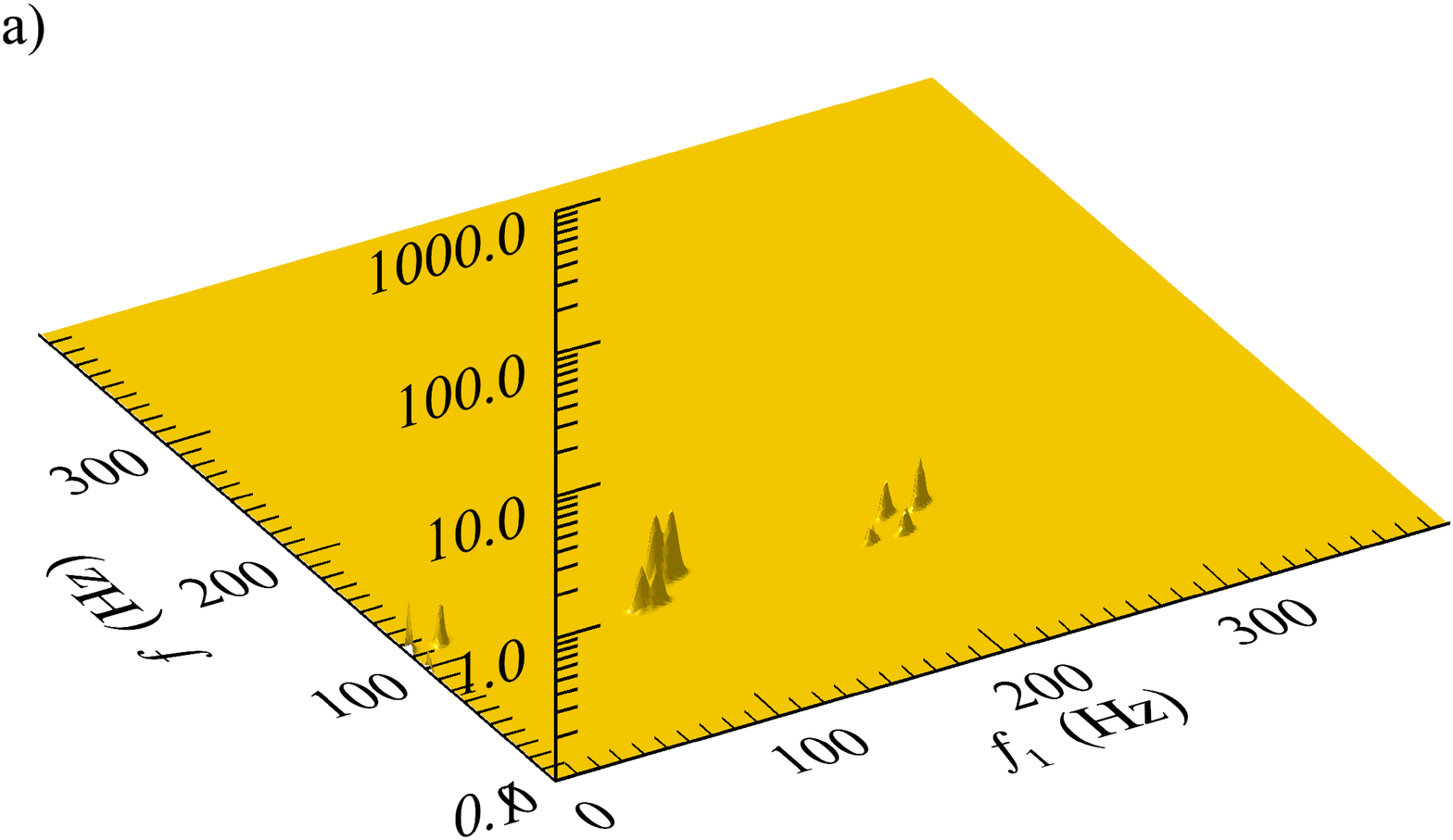}
\end{minipage}%
\begin{minipage}{.5\textwidth}
  \centering
  \includegraphics[width=0.98\linewidth]{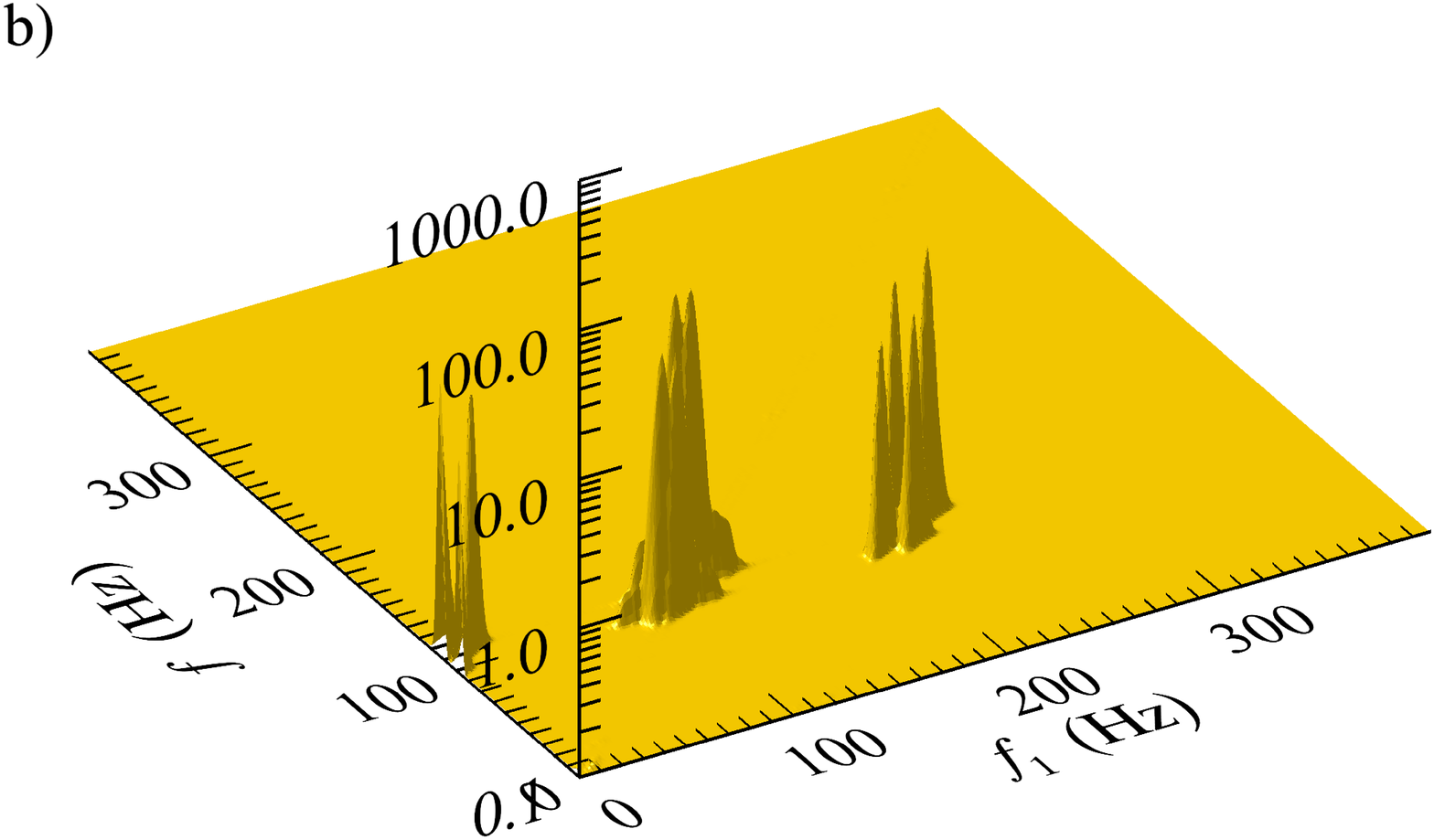}
\end{minipage}
\begin{minipage}{0.5\textwidth}
  \centering
  \includegraphics[width=0.99\linewidth]{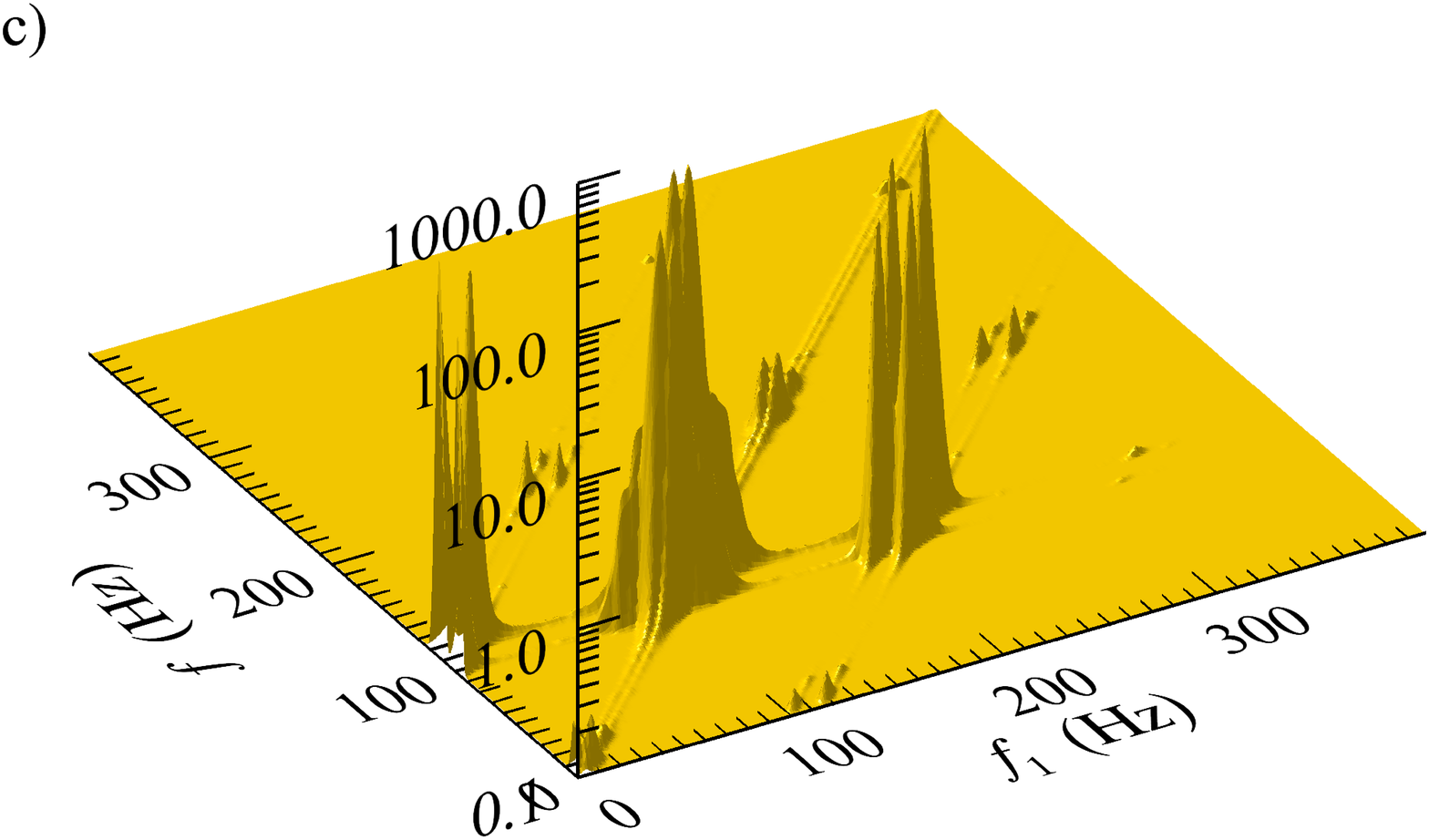}
\end{minipage}%
\begin{minipage}{.5\textwidth}
  \centering
  \includegraphics[width=0.98\linewidth]{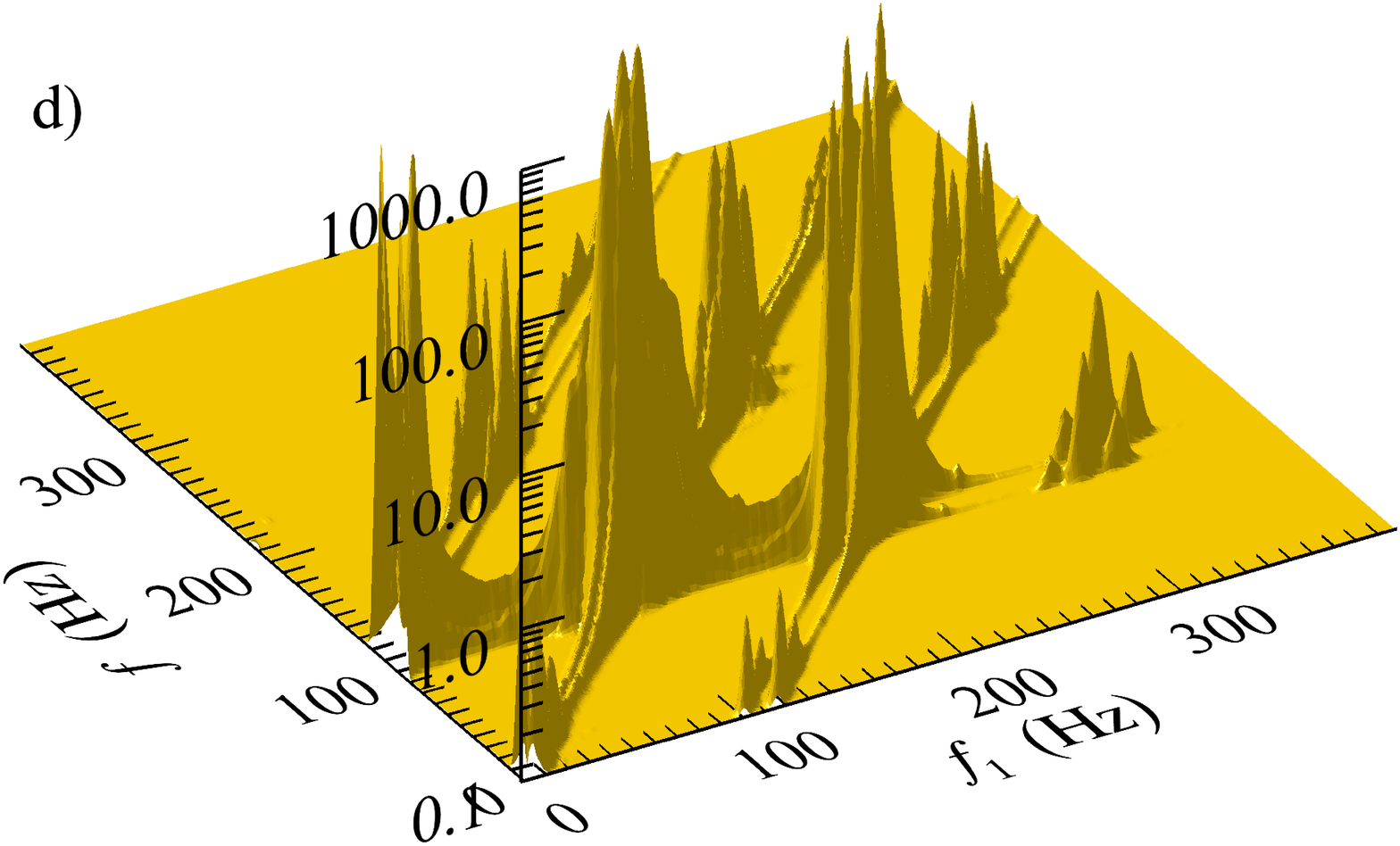}
\end{minipage}
    \caption{Time development of iterative additions to dual-rod triad interactions (first quadrant) for the measured two-rod signal, 100 records averaged. (a) 10 iterations, (b) 20 iterations, (c) 30 iterations and (d) 40 iterations.}
    \label{fig:19}
\end{figure*}

Figure~\ref{fig:19} shows corresponding phase matched triad interactions averaged over 100 records. Again, we see the development of the main group of frequency components, but as in Figure~\ref{fig:18}, the detailed structure is difficult to discern due to unsteadiness in the vortex frequencies and the surrounding flow.

\section{\label{sec:Conclusion}Conclusions}

The development of power spectra measured downstream from one and two vortex shedding rods in a low intensity uniform flow have been presented. We may view the vortices as large oscillating modes injected into the flow, and we can measure the development of higher frequency modes as a result of a sequence of triad interactions caused by the second order nonlinearity in the Navier-Stokes equation between the different modes as the fluid is convected downstream in the flow. We then used the iterative Navier-Stokes program to compute the development of the power spectrum of an initial velocity record as a function of time, both with some computer generated single or two-frequency signals and with the hot-wire signal measured near the rods as input signal. The NSE program allows us to study the individual contributions to the power spectrum as a function of time, i.e. with increasing number of computational iterations. This gives unique insight into the dynamics of the individual triad interactions, both in case of simulated modes and for an actual measured velocity signal as input to the program. As we get good agreement between the measured and the computed power spectra, we assume that we with confidence can use the computer program to study the interactions between all the modes present. These interactions are available during the computations, both as regards the actual interacting frequencies and the strengths of these interactions as well as the time constants in the interactions. From both the power spectrum plots and the surface plots of the triad interactions, we see clearly the effect of the unique form of the nonlinear term in the Navier-Stokes equation, which forces kinetic energy \textit{on average} to move to higher and higher frequencies, which is the true reason for the so-called energy cascade.

\section*{Acknowledgements}

CMV acknowledges financial support from the European Research council: This project has received funding from the European Research Council (ERC) under the European Unions Horizon 2020 research and innovation program (grant agreement No 803419). 

MR and PB acknowledge financial support from the Poul Due Jensen Foundation: Financial support from the Poul Due Jensen Foundation (Grundfos Foundation) for this research is gratefully acknowledged.

\section*{Declaration of interest}
The authors report no conflicts of interest. 

\section*{Data availability}

The data that forms the basis of this study is available from the corresponding author upon reasonable request. 

\nocite{*}
\section*{References}
\bibliography{aipsamp}

\end{document}